\documentclass[aps, english, notitlepage,superscriptaddress]{revtex4-1}

\usepackage{mathptmx}
\usepackage[T1]{fontenc}
\usepackage{amssymb}
\usepackage{graphicx}
\usepackage{amsmath,color}
\usepackage{natbib,hyperref}
\usepackage{multirow,bigdelim}
\usepackage{amsfonts, bm}

\usepackage{lipsum}
\usepackage{gensymb}
\usepackage{xhfill}
\usepackage{enumitem}
\usepackage[skins]{tcolorbox}
\usepackage{varwidth}

\DeclareMathAlphabet{\mathcal}{OMS}{cmsy}{m}{n}

\begin{document}

\title{Data-driven path collective variables}

\author{Arthur France-Lanord}
\email{arthur.france-lanord@cnrs.fr}
\affiliation{Sorbonne Université, Institut des Sciences du Calcul et des Données, ISCD, F-75005 Paris, France}
\affiliation{Sorbonne Université, Muséum National d’Histoire Naturelle, UMR CNRS 7590, Institut de Minéralogie, de Physique des Matériaux et de Cosmochimie, IMPMC, F-75005 Paris, France}
\author{Hadrien Vroylandt}
\affiliation{Sorbonne Université, Institut des Sciences du Calcul et des Données, ISCD, F-75005 Paris, France}
\author{Mathieu Salanne}
\affiliation{Physicochimie des \'Electrolytes et Nanosyst\`emes Interfaciaux, Sorbonne Universit\'e, CNRS, 4 Place Jussieu F-75005 Paris, France}
\affiliation{Institut Universitaire de France (IUF), 75231 Paris, France}
\author{Benjamin Rotenberg}
\affiliation{Physicochimie des \'Electrolytes et Nanosyst\`emes Interfaciaux, Sorbonne Universit\'e, CNRS, 4 Place Jussieu F-75005 Paris, France}
\author{A. Marco Saitta}
\affiliation{Sorbonne Université, Muséum National d’Histoire Naturelle, UMR CNRS 7590, Institut de Minéralogie, de Physique des Matériaux et de Cosmochimie, IMPMC, F-75005 Paris, France}
\author{Fabio Pietrucci}
\email{fabio.pietrucci@sorbonne-universite.fr}
\affiliation{Sorbonne Université, Muséum National d’Histoire Naturelle, UMR CNRS 7590, Institut de Minéralogie, de Physique des Matériaux et de Cosmochimie, IMPMC, F-75005 Paris, France}

\begin{abstract}
Identifying optimal collective variables to model transformations, using atomic-scale simulations, is a long-standing challenge. 
We propose a new method for the generation, optimization, and comparison of collective variables, which can be thought of as a data-driven generalization of the path collective variable concept. It consists in a kernel ridge regression of the committor probability, which encodes a transformation's progress. The resulting collective variable is one-dimensional, interpretable, and differentiable, making it appropriate for enhanced sampling simulations requiring biasing. We demonstrate the validity of the method on two different applications: a precipitation model, and the association of Li$^+$ and F$^-$ in water. For the former, we show that global descriptors such as the permutation invariant vector allow to reach an accuracy far from the one achieved \textit{via} simpler, more intuitive variables. For the latter, we show that information correlated with the transformation mechanism is contained in the first solvation shell only, and that inertial effects prevent the derivation of optimal collective variables from the atomic positions only. 
\end{abstract}

\maketitle

\section{Introduction}

Accurately modeling how matter transforms at the atomic scale is key in understanding many fundamental physical phenomena, ranging from phase transitions to chemical reactions. An ideal approach to studying such phenomena {\it in silico} would consist in simply observing the time evolution of an atomistic system, at a given set of thermodynamic conditions; {\it i.e.} an equilibrium molecular dynamics simulation. However, the free energy barriers associated with transformations are usually several times larger than the thermal energy $k_B T$, so that naturally observing and sufficiently sampling these rare phenomena is simply unachievable. Researchers have therefore developed advanced methods to bias the dynamics of a given system, allowing to effectively lower selected barriers of interest, and to enhance the sampling of rare events over the course of a molecular dynamics simulation. This set of methods\cite{henin2022} has been applied to many different physical problems, including the phase diagram of water\cite{pipolo2017} and its putative liquid-liquid phase transition\cite{jedrecy2023,gartner2022}, ice nucleation\cite{lupi2017}, protein folding dynamics\cite{shea2001}, and chemical reactions governing prebiotic chemistry\cite{saitta2014}.

The difficulty and subtlety associated with enhanced sampling methods comes from the high dimensionality ($3N$, where $N$ is the number of particles) of configuration space, {\it i.e.} the space spanning all possible atomic configurations. Since it is virtually impossible to directly identify relevant barriers in such a high-dimensional space, the solution is to select {\it collective variables} (CVs), expressed as a function of the system's degrees of freedom. By doing so, the configuration space is projected onto one of much lower dimension. Good CVs should allow to distinguish basins of metastability, as well as transition state ensembles. The way they are chosen affects many crucial elements: i) the observed transformation mechanism in the context of enhanced sampling; ii) CVs are often used as a means of analyzing a reaction mechanism, be it observed from biased or unbiased dynamics. A poorly selected CV can therefore lead to a biased or incomplete rationalization of the transformation mechanism, even in the context of unbiased simulations; iii) free energies of activation, $\Delta F^{\ddag}$, are sometimes extracted from gauge-invariant free energy profiles\cite{hartmann2011,bal2020,dietschreit2022} along collective variables, and used to interpret the kinetics of the system of interest. Using a sub-optimal CV will lead, in this context, to 
an underestimated $\Delta F^{\ddag}$ for the transformation under scrutiny\cite{bal2020}; iv) while in the context of the Eyring-Polanyi equation kinetic rates are independent of the choice of CV, thus a poorly selected one will have an associated transmission coefficient far from unity; v) sub-optimal CVs hamper the efficiency of enhanced sampling methods based on bias forces, since the latter are partly wasted along degrees-of-freedom not related to the transformation, leading to the unphysical overestimation of the free energy barrier height~\cite{palacio22rates}; vi) on the contrary, sub-optimal CVs result also in an under-estimation of the free-energy barrier from the viewpoint of the ideal, exact calculation based on marginalizing the canonical probability density (see, \textit{e.g.}, Eq.~\ref{e:fe} here below), due to overlap between metastable basins~\cite{jungblut2016}. The latter effect has been exploited in a CV optimization approach based on rate minimization~\cite{mouaffac2023}, 
and concerns, more generally, Langevin models describing the  high-dimensional dynamics projected on CVs, constructed from statistical inference on unbiased MD trajectories\cite{vroylandt2022,girardier2023}.

Selecting a CV on solid grounds is therefore an active field of research, with many methods developed in the last decades, following different approaches. Identifying slow modes is one: this has been achieved using diffusion maps\cite{nadler2006,coifman2008,rohrdanz2011,rydzewski2022}, variational approaches based on the overdamped Langevin equation\cite{mouaffac2023,zhang2016} or Markov state models\cite{noe2013} -- combined with independent component analysis\cite{perez2013,molgedey1994} and deep learning\cite{mardt2018} --, or maximizing the spectral gap of a transition matrix\cite{tiwary2016,rydzewski2023}. Another set of methods relies on (auto)encoders\cite{hernandez2018,ribeiro2018} and the information bottleneck principle\cite{wang2019,wang2021}. A few techniques\cite{mendels2018,sultan2018,bonati2020} analyze fluctuations in the metastable states to extract optimal CVs. 

Finally, a large body of work has been dedicated to deriving optimal collective variables based on the committor probability $p(\text{B}|\mathbf{X})$, \textit{i.e.} in a system showing two metastable states A and B, the probability of reaching basin B first, conditioned on an initial point $\mathbf{X}$ in configuration space. The committor is, by construction, the reaction coordinate of the transformation under scrutiny, as it encodes the entire dynamical process of transforming from one metastable state into the other\cite{ma2005}. Depending on how optimality is defined, it may or may not be an optimal collective variable: for instance, in the presence of large free energy barriers, the committor's strong non-linearity near basins\cite{lechner2010} is impractical to handle. Nevertheless, an intuitive approach is therefore to perform a regression of the committor using a flexible basis of collective variables; this has been pioneered by Ma and Dinner\cite{ma2005} using neural networks and a genetic algorithm. Later, Peters and Trout\cite{peters2006} devised a fitting approach based on transition path sampling\cite{bolhuis2002} data and the transition path probability $p(\text{TP}|\mathbf{X})$, which is analytically related to the committor under the assumption of diffusive dynamics. This was later expanded to obtain CVs with maximal transmission coefficients\cite{peters2012}, or using cross-entropy minimization and regularization\cite{mori2020}. Recently, this approach has been recast in a reinforcement learning framework to allow the identification of CVs over the course of transition path sampling simulations\cite{jung2023,jung2023b}, including an extension to extract free energies and rates\cite{Lazzeri23}. Committor-based methods have also been combined with slow process identification\cite{chen2023}. 

In this article, we propose a committor regression technique for CV selection based on kernel ridge regression, which as we show can be thought of as a data-driven generalization of path collective variables\cite{branduardi2007}. The resulting optimized collective variable is one-dimensional and differentiable, making it appropriate for biased simulations. It is interpretable, provided that the CV subspace it is based on is interpretable as well. 

This paper is structured as follows: in Sec. \ref{s:theo}, we introduce the method and discuss its connection to path collective variables. In Sec. \ref{s:rmb}, we apply the method to a two-dimensional toy potential for which we can compute the committor exactly. We then turn to more realistic examples, with a precipitation phenomenon using Lennard-Jones particles (Sec. \ref{s:lj}), and the association of small, monoatomic ions in water (Sec. \ref{s:lif}). Finally, we conclude and provide an outlook regarding possible extensions to the method, in Sec. \ref{s:c}.

\section{From standard to data-driven path collective variables}\label{s:theo}

\begin{figure}[h!]
\centering
\includegraphics[width=.5\textwidth]{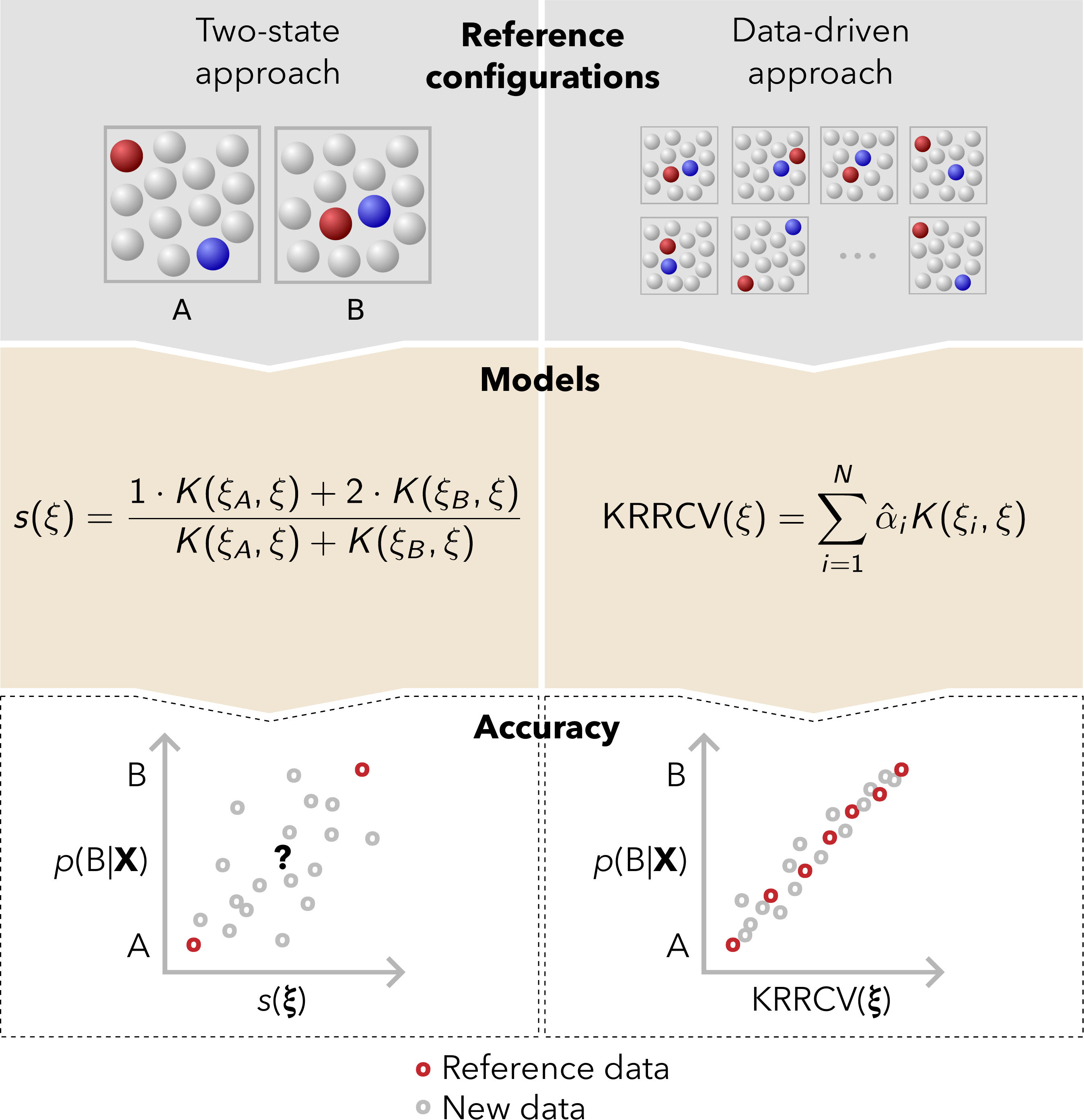}
\caption{Standard, or two-state (left) and data-driven (right) path collective variables. Consider a system made of a red particle, a blue particle, and several gray particles. The red and blue particles can be in two metastable states: dissociated (A) or associated (B). Left: two reference states describing metastable states A and B are fed to the path CV. This will lead to a good description of configurations close to or in either metastable states, but the behavior outside of the basins is difficult to predict. This can be seen when comparing the committor $p(\text{B}|\mathbf{X})$ and $s(\xi)$: configurations far from the basins do not have close by reference states. Right: in a data-driven approach, a much greater number of references is considered, describing the full reaction process, from state A to state B. The committor has to be computed for each reference, and is fed to the path collective variable through kernel ridge regression. Agreement with the true committor function is improved as reference states are distributed over the whole reaction pathway. }
\label{fig1}
\end{figure}

Path collective variables\cite{branduardi2007,pietrucci2015} describe the progress along a reaction pathway connecting metastable states. The main concept, summarized in the top part of Fig. \ref{fig1}, is to define reference states, and to compute, for a given atomic configuration, the similarity to these reference states using a kernel function $K(\xi_i, \xi_j)$. The similarity is not computed on the raw atomic coordinates, but rather on a vector of collective variables of arbitrary dimension, $\xi$. For instance, coordination numbers of a specific set of atoms are often used for chemical reactions\cite{afl2021,pietrucci2015}. Path CVs therefore involve two projections: from the atomic positions $\mathbf{X}$ to intermediate CVs $\xi$ (which live in what we call the \textit{CV subspace}), and from $\xi$ to the one-dimensional path CV denoted $s(\xi)$:

\begin{equation}\label{e:pcv}
s(\xi) = \frac{\sum_{i=1}^N iK(\xi_i, \xi)}{\sum_{j=1}^N K(\xi_j, \xi)}. 
\end{equation}

A typical choice for the kernel is $K = \text{exp}\left( -\lambda \| \xi_i - \xi \|_2^2 \right)$, \textit{i.e.} the radial basis function kernel, where $\lambda$ is a bandwidth parameter, and $\| \xi_i - \xi \|_2^2$ the squared $l^2$-norm. Conceptually, $s$ describes the progress along the polygonal chain in $\xi$-space connecting all references. An additional quantity, $z$, is usually defined to express the distance to this pathway\cite{branduardi2007}. Traditionally, $\mathcal{O}(10^1)$ equidistant references in CV subspace are selected along the pathway using nudged-elastic-band-like algorithms\cite{branduardi2007,saladino2012}; they can also be optimized during enhanced sampling simulations with an adaptive approach\cite{perez2019}. Later, a simpler and more explorative approach 
proposed to set

the number of reference states to the number of identified metastable states of interest\cite{pietrucci2015,pipolo2017}. For instance, in the association/dissociation problem considered in Fig. \ref{fig1}, the number of states is set to two. In any case, it is then assumed that the path CV will be able to interpolate correctly for configurations halfway between both states. A more important assumption is that the selected CV subspace $\xi$ is able to resolve the transformation. At this stage, both are uncontrolled approximations that should be thoroughly verified, as a good CV should capture the transition state ensemble, whose configurations do not belong to either states. 

It is therefore natural to consider including many reference states in the path CV definition. As illustrated in Fig. \ref{fig1}, this can only improve CV quality along the whole reaction path, \textit{i.e.} the pathway, in CV subpace, that connects metastable states. However, a measure of the reaction progress is now needed. Indeed, $s$ can be seen as a weighted average; for two reference states, $i$ is simply set to 1 and 2. A configuration close to basin $A$ will therefore have a $s$ close to 1. If one wants to include reference states outside of metastable basins, it is therefore mandatory to use a measure of the progress along the reaction path, with some methods implemented to evenly space references in the CV subspace\cite{hovan2018,magrino2022}. 

One way to evaluate the quality of a collective variable is to compare it to the committor probability, which is widely considered the optimal CV\cite{peters2006} for a system of two metastable states. For a configuration already in basin $A$ $p(\text{B}|\mathbf{X}\in \text{A})=0$, and $p(\text{B}|\mathbf{X}\in \text{B})=1$ for one in state $B$. At the transition state, $p(\text{B}|\mathbf{X}\in \text{TS})=0.5$. Unfortunately, for realistic -- \textit{i.e.} high-dimensional -- systems, computing the committor for a given configuration is computationally expensive: one has to run a series of short independent simulations by randomly drawing initial velocities from the Maxwell-Boltzmann distribution. These simulations end when the system reaches either basins, and to reach a satisfying accuracy on the committor, it is necessary to generate hundreds-to-thousands of short trajectories 
when the committor value is not too close to zero or one\cite{peters2006error}. In the presence of high barriers the transition mechanism largely involves configurations with very tiny deviations of the committor from zero or one, requiring more sophisticated estimation approaches \cite{lechner2010}.

In addition, the committor, estimated numerically in this way, is not differentiable, which is problematic for biased simulations. While it is therefore nearly impossible to use the committor directly as a CV, one can compute values for an ensemble of reference configurations, and use these values as weights in the path CV. This is the main idea behind the present data-driven path collective variables. 

Interestingly, the expression of $s$, which is shown in Fig. \ref{fig1}, is that of a kernel regression (or kernel smoothing) estimator. Kernel regression is a class of non-parametric methods for non-linear regression. Since kernel regression improves with increasing amounts of data, it is even more natural to consider using many reference states. Surprisingly, the connection to kernel regression was not made in the original article on path CVs\cite{branduardi2007}, or in any other subsequent work. In greater detail, $s$ is the Nadaraya-Watson estimator\cite{nadaraya1964,watson1964}, which is the zeroth-order approximation to the local polynomial estimator, also called the "local constant" approximation. We switch to a global method in the form of kernel ridge regression (KRR), \textit{i.e.} the combination of linear least squares with $l^2$-norm regularization and the kernel trick. The KRR estimator is: 

\begin{equation}
\hat{f}(\xi) = \sum_{i=1}^N \alpha_i K(\xi_i,\xi),
\end{equation}

optimal parameters $\hat{\bm{\alpha}} = [\hat{\alpha}_1, \dots, \hat{\alpha}_n]^T$ are obtained by solving the following linear system: 

\begin{equation}\label{e:akrr}
\hat{\bm{\alpha}} = \arg\min_{\bm{\alpha}\in \mathbb{R}^N} \| y - K \bm{\alpha} \|_2^2 + \lambda \bm{\alpha}^T\bm{\alpha} = (K_{NN} + \lambda \mathbb{I}_n)^{-1} y,
\end{equation}

where we introduced the kernel matrix of all pairs of data $K_{NN}$, the target vector $y = [p(\text{B}|\mathbf{X}_1),\dots, p(\text{B}|\mathbf{X}_N)]^T$, and the regularization parameter $\lambda$. In addition, we introduce a bandwidth matrix $\Sigma$ such that our kernel becomes: 

\begin{equation}
K(\xi_i, \xi) = \exp{\left(-(\xi_i - \xi)^T \Sigma^{-1} (\xi_i - \xi)\right)},
\end{equation}

allowing more flexibility in reproducing the target data. In the following, $\Sigma$ is always diagonal and reduces to a bandwidth vector $\bm{\sigma} = [\sigma_1, \dots, \sigma_d]^T$, for $\xi \in \mathbb{R}^d$. In this way, each component in $\xi$ has a corresponding weight which encodes its importance: a low $\sigma$ means that the corresponding CV component is highly correlated to the committor. $\xi$ can be composed of simple, intuitive collective variables, but can also include high-dimensional, abstract representations such as commonly used local descriptors for machine-learned interatomic potentials (\textit{e.g.} atom-centered symmetry functions\cite{behler2007} and the smooth overlap of atomic positions (SOAP)\cite{bartok2013}), or global variants such as the permutation-invariant vector (PIV)\cite{gallet2013,pipolo2017}. Optimal values for $\bm{\sigma}$ and the regularization parameter $\lambda$ are obtained through optimization, by minimizing a loss function based on a training set, distinct from the set of configurations used in KRR (the reference set). 

\section{The rugged Müller-Brown potential}\label{s:rmb}

\begin{figure}[h!]
\centering
\includegraphics[width=\textwidth]{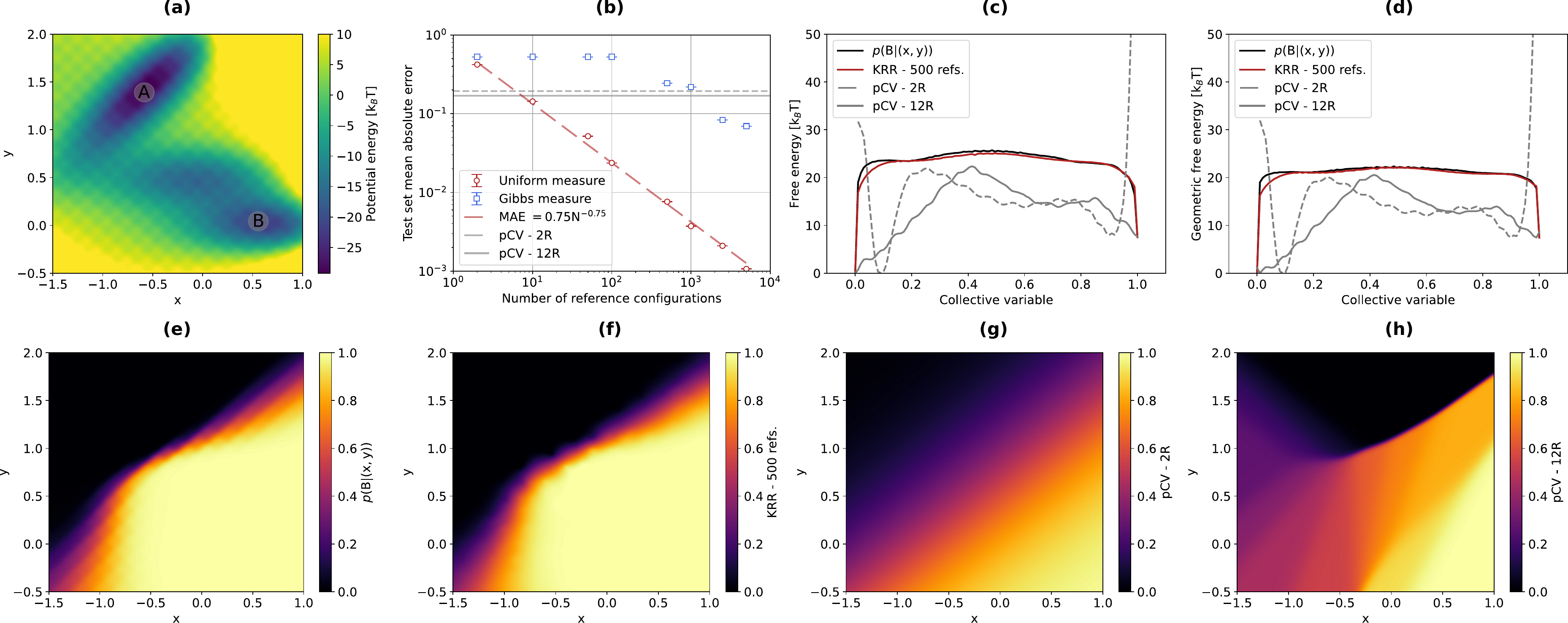}
\caption{Path collective variables for the rugged Müller-Brown potential. (a) The potential energy surface. Shaded circles correspond to the metastable states definition used for the evaluation of the committor, (b) test set mean absolute error for different path collective variables as a function of the number of reference configurations included. Red circles and blue squares respectively correspond to models built on data sampled from the uniform and Gibbs measures. Error bars correspond to 95\% confidence intervals of a distribution of predictions spanning ten different reference and training sets. The red broken line is a power law fit. The full and broken gray lines respectively correspond to the "12R" and "2R" path collective variables, (c) canonical and (d) geometric free energy profiles along various path collective variables, compared to the committor (small wiggles are due to the numerical integration and marginalization), (e-h) collective variables plotted in configuration space: (e) the committor as obtained from the backward Kolmogorov equation, (f) the kernel ridge regression based on 500 references, (g) the "2R" and (g) "12R" path collective variables.}
\label{fig2}
\end{figure}

As a starting point, we select the two-dimensional rugged Müller-Brown\cite{lai2018,li2019} (rMB) analytical potential. Assuming overdamped Langevin dynamics, we can calculate the committor probability over the whole configuration space, as explained in Appendix \ref{a:fe}. This allows us to assess the quality of various regression models without projection errors, by considering the configuration space as the CV subspace. The rMB potential takes the form of the traditional Müller-Brown potential\cite{muller1979} with added ruggedness: 

\begin{equation}\label{e:rmb}
\begin{split}
V(x,y) = & \sum_{i=1}^4 D_i \exp \left[ a_i (x - X_i)^2 + b_i (x - X_i)(y - Y_i) \right. \\ 
& \left. + c_i (y - Y_i)^2 \right] + \gamma \sin (2k\pi x) \sin (2k\pi y),
\end{split}
\end{equation}

where $D_{1...4} = [-400, -200, -340, 30]$, $a_{1...4} = [-1, -1, -6.5, 0.7]$, $b_{1...4} = [0, 0, 11, 0.6]$, $c_{1...4} = [-10, -10, -6.5, 0.7]$, $X_{1...4} = [1, 0, -0.5, -1]$, $Y_{1...4} = [0, 0.5, 1.5, 1]$, $\gamma = 9$, and $k = 5$. We define metastable states $A$ and $B$, shown in Fig. \ref{fig2}(a), as circles of radius 0.1 and centered respectively at $(-0.58,1.39)$ and $(0.55,0.05)$. 

To construct datasets for KRR CVs (\textit{i.e.} a reference set and a training set), we sample $p(\text{B}|(x,y))$ values from the uniform measure or from the Gibbs measure on the configuration space. Sampling configurations in high-dimensional space -- especially in the presence of potential energy barriers -- is a difficult task, which is why we consider two radically different sampling strategies. As we will see in the next sections, for realistic systems, enhanced sampling strategies are leveraged. The test set is always the same for all models, consisting of 4000 data points sampled randomly. The detailed procedure for model optimization is reported in Appendix \ref{a:opt}. We assess the accuracy of the various path collective variables by comparing their mean absolute error (MAE) on the test set (Fig. \ref{fig2}(b)), their value over configuration space \ref{fig2}(d,e)), and free energy profiles \ref{fig2}(f)). Free energy profiles are obtained by evaluating the collective variable's marginal probability density over configuration space: 

\begin{equation}\label{e:fe}
F(\xi) = - \beta^{-1} \ln \iint e^{-\beta V(x,y)} \delta \left( \xi(x,y) - \xi \right) \mathrm{d}x \mathrm{d}y, 
\end{equation}

where $\beta=0.1$. We also report a "geometric" free energy profile, $F_g(\xi)$, distinct from the canonical definition of equation \ref{e:fe}. The canonical free energy profile is not left invariant by a gauge transformation of a CV, which makes it impossible to compare different choices of $\xi$, even if they contain the same amount of information with respect to the reaction\cite{bal2020,dietschreit2022}. The geometric definition incorporates this gauge invariance, which, as we will see later on, will be useful to compare quantities such as free energies of activation for different CVs. The geometric free energy profile can be derived through the application of the coarea formula, as has been discussed many times before\cite{weinan2004,hartmann2011}: 

\begin{equation}\label{fe_g}
F_g(\xi) = - \beta^{-1} \ln \iint e^{-\beta V(x,y)} \left| \nabla \xi(x,y) \right| \delta \left( \xi(x,y) - \xi \right) \mathrm{d}x \mathrm{d}y.
\end{equation}

Both profiles are related in the following way: 

\begin{equation}\label{fe_g_fe}
e^{-\beta F_g(\xi)} = \left\langle \left| \nabla \xi(x,y) \right| \right\rangle_{\xi} e^{-\beta F(\xi)}, 
\end{equation}

where $\left\langle . \right\rangle$ is an ensemble average. We also construct two variants of traditional path CVs: the first one ("2R") containing two references -- one in each basin --, and a more advanced version ("12R"), containing twelve references equally spaced along the minimum energy path, including the exact same two as for the "2R" version. The latter one is assumed to be more advanced than the former since the minimum energy path is usually not available \textit{a priori} for realistic, high-dimensional settings. References are respectively labeled $1,2$ and $1,\ldots,12$; to compare with the committor, the path CVs are scaled to the $[0,1]$ interval. The scalar bandwidth parameter $\lambda$ is set so that the distance between the first and second references times $\lambda$ is equal to 2.3, a common practice leading to smooth free-energy landscapes, typically with a good separation of reactants, products and transition states\cite{branduardi2007,pietrucci2015}. 

Results are shown on Fig. \ref{fig2}, with the global ability in reproducing the committor of each method reported in Fig. \ref{fig2}(b). The largest MAE attainable is roughly 0.5, since the test set is composed of values in the $[0,1]$ interval with mean roughly equal to 0.5. As one would expect, the test set MAE decreases with increasing number of references. However, sampling from the Gibbs measure is demonstrated to be largely inefficient, which obviously shows the importance of enhanced sampling for realistic systems. Sampling from the Gibbs measure leads to selecting points mostly in the metastable states, where the committor is constant, while it rapidly varies near the separatrix. This is even more critical with a low temperature or large energy barriers. 

Canonical path CVs, while they do not aim at reproducing the committor, perform relatively decently, with test set MAEs respectively equal to 0.194 and 0.169 for the "2R" and "12R" variants (see also Fig. \ref{fig2}(g,h)). Using only two references, or aligned references in the CV subspace, leads to constant path CVs along the direction normal to the line connecting the references. Here, as can be seen in \ref{fig2}(e), there are non-negligible variations of the committor in directions normal to the line connecting both metastable states. In addition, with no measure of the reaction progress, one can only label references in a linear fashion. 

As mentioned before, the true committor varies non-linearly, depending on the underlying potential energy surface and the temperature. This highlights the importance of extracting such information, possibly by computing committor values. We note that other methods are used for labeling references, for instance based on transition path sampling data \cite{magrino2022}. With a "reasonable" number of 500 references (reasonable because it can be rather easily achieved with moderate computational cost for the realistic examples reported in the next sections), one can achieve excellent, yet still improvable, agreement with the true committor, with a test set MAE smaller than $10^{-2}$. This also translates into a qualitative agreement across configuration space (Fig. \ref{fig2}(e,f)), and quantitative agreement of free energy profiles (Fig. \ref{fig2}(c,d)). 

Features of the free energy profiles along path collective variables such as the free energy of activation are in good agreement with the committor. Depending on the system, this might not always be the case: in Appendix \ref{a:3w}, we show results for a three wells model potential for which simple path collective variables do not perform as well. Realistic, high-dimensional systems typically show many shallow intermediate states. Nevertheless, these results demonstrate the strength of path collective variables. Of course, this is valid only in the limit where the CV subspace is the configuration space itself, \textit{i.e.} when there is no loss of information due to a projection. 

\begin{figure}[h!]
\centering
\includegraphics[width=.5\textwidth]{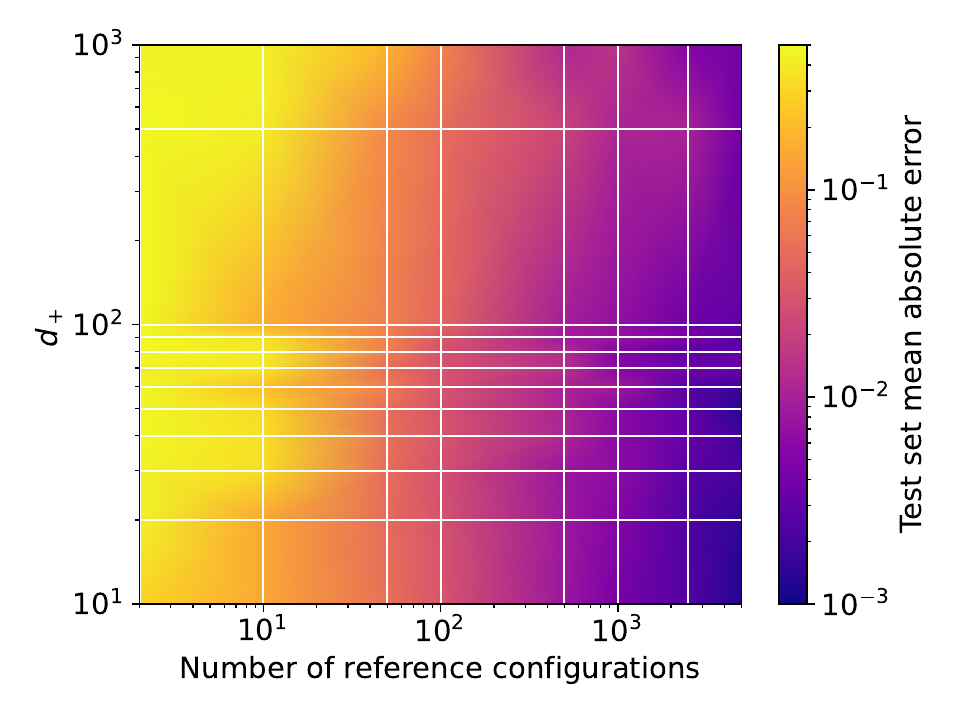}
\caption{Test set mean absolute error for KRR models trained using different number of references and parasitic dimensions. The error is interpolated over data points represented by the intersections of the white lines (including $N=2$ and $d_+=10$).}
\label{fig3}
\end{figure}

To assess the robustness of the KRR approach, we add parasitic dimensions to the potential, uncorrelated to the committor: 

\begin{equation}\label{e:potdplus}
V(x,y,\mathbf{z}) = V(x,y) + \sum_{j=1}^{d_+} z_j^2,
\end{equation}

in which we vary $d_+$ from 10 to 1000. The test set mean absolute error as a function of $d_+$ and the number of references is presented in Fig. \ref{fig3}. As one can see, variations along $d_+$ are relatively negligible. This mostly demonstrates the robustness of the bandwidth optimization approach, since the problem here is finding directions correlated with the committor in a high-dimensional space. Finally, as shown in Appendix \ref{a:5d}, we can also embed the two-dimensional potential in a five-dimensional space using non-linear transformations leading to large degeneracies, and still recover a low test set MAE. This is perhaps expected, since kernel ridge regression models can learn non-linear features. 

\section{Precipitation of Lennard-Jones particles}\label{s:lj}

We now turn to a more realistic system involving Lennard-Jones interactions, aimed at capturing the onset of a precipitation process. Precipitation is a critical phenomenon in inorganic or organic chemistry, as well as in biochemistry and metallurgy. For instance, in Li-ion batteries, decomposition products precipitate to form a solid interphase between a solid electrode and the liquid electrolyte. Here 20 particles represent the precipitating species, and 4086 smaller particles represent the solvent. The solute species interact strongly with one another ($\epsilon_{11} = 3.1$, $\sigma_{11} = 1.2$), while the solvent-solvent and solute-solvent interactions are weaker ($\epsilon_{12} = \epsilon_{22} = 1.0$, $\sigma_{12} = \sigma_{22} = 1.0$). The interaction cutoff is set to $6 \sigma$. All masses are set to 1. These parameters are chosen such that two metastable states separated by a free energy barrier exist: an associated state, where solute particles clump together to form an aggregate, and a dissociated state, where solvent particles coordinate solute particles. The integration time step ($\delta t$) is set to $2\cdot10^{-3}$, and all simulations are run at $T=1$, $p=1$, using the LAMMPS program\cite{thompson2022} (version 7 Aug 2019), with PLUMED\cite{bonomi2019,tribello2014} (version 2.5.3) as an add-on to compute collective variables and perform enhanced sampling. 

Our sampling strategy is the following: 1) converging free energy profiles from unbiased molecular dynamics, 2) sampling the putative transition state ensemble using a simple collective variable, 3) sampling close to the true transition state ensemble using transition path sampling. The committor is numerically estimated for configurations extracted from steps 2) and 3); these pairs of structure and committor constitute the datasets we then use to optimize and validate KRRCV models. All computational details regarding these steps are presented in Appendix \ref{a:lj}.

We start by obtaining free energy profiles along two simple one-dimensional collective variables, which we select to be properties of the largest aggregate. We use a depth-first search (DFS) clustering algorithm to identify the largest aggregate in a given configuration\cite{tribello2017}, using a distance-based criterion to assess if particles are aggregated or not. Our two collective variables correspond to the number of particles in the largest cluster ($\Sigma N$), and of the sum of solute coordination numbers over solute particles in the largest cluster ($\Sigma C$). Profiles are obtained from long unbiased molecular dynamics in the canonical ($nVT$) ensemble. 

\begin{figure}[h!]
\centering
\includegraphics[width=\textwidth]{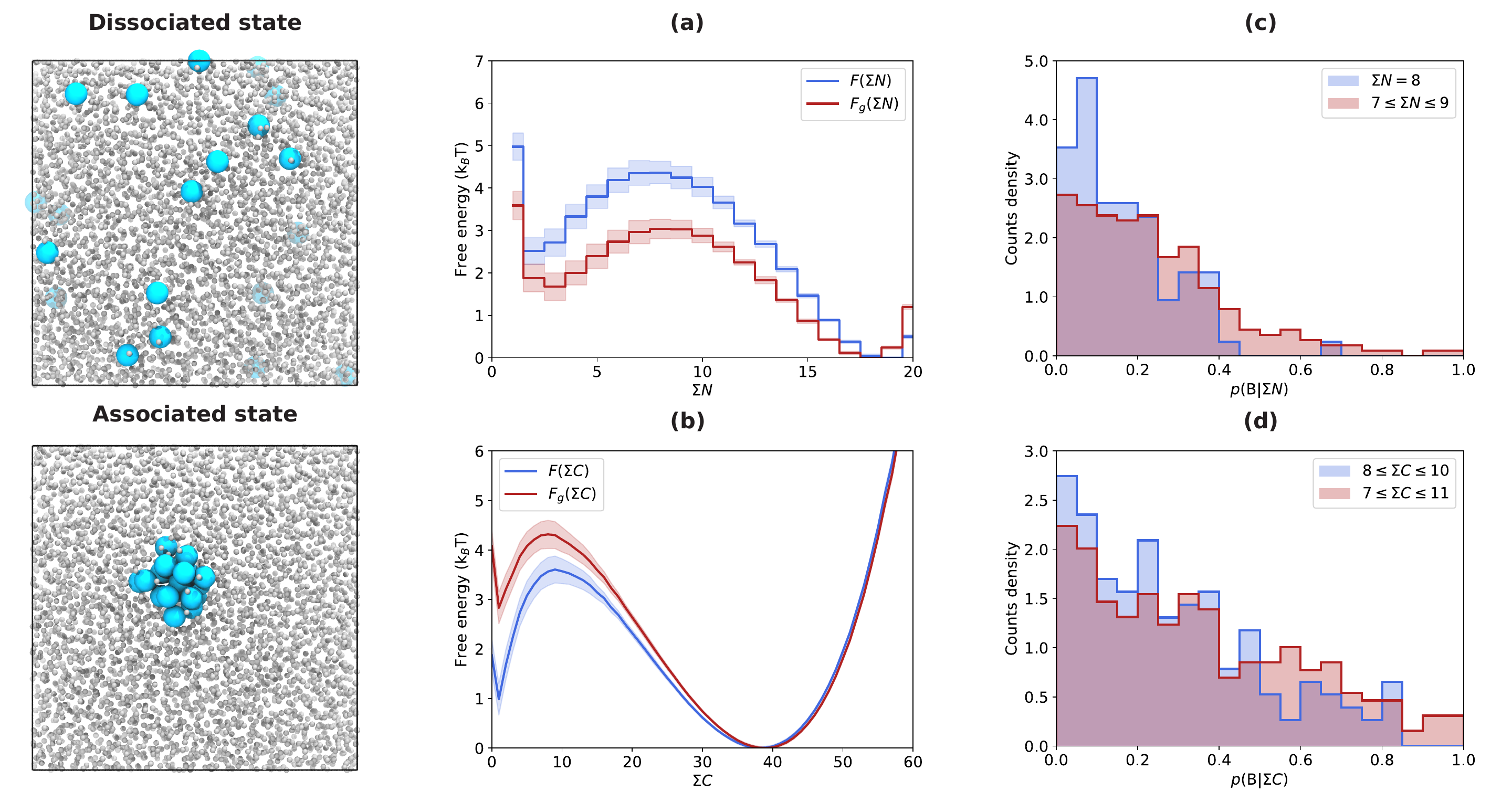}
\caption{Lennard-Jones model of a precipitation phenomenon. Left panels: system snapshots in the dissociated and associated states. Large cyan and small gray spheres respectively represent solute and solvent particles. (a,b) Free energy profiles along collective variables $\Sigma N$ and $\Sigma C$. (c,d) distributions of committor values for configurations at the putative transition state ensemble of $\Sigma N$ and $\Sigma C$. }
\label{fig4}
\end{figure}

Free energy profiles are reported in Fig. \ref{fig4}(a,b). As can be seen, with both $\Sigma N$ and $\Sigma C$, the associated state is considered more thermodynamically stable than the dissociated one; in addition, both states are separated by a barrier of a few $k_BT$. 

Enhanced sampling involving biasing along $\Sigma N$ is not achievable as it is a discontinuous quantity; we therefore perform umbrella sampling along $\Sigma C$\cite{tribello2017}, from which we extract 500 configurations. We also perform transition path sampling simulations, using the flexible-length aimless shooting algorithm\cite{peters2006,mullen2015}. We initialize such simulations by generating 100 reactive trajectories connecting both basins. The basin definition is the following: $A: \Sigma N < 5$, $B: \Sigma N > 14$. From these initial reactive paths, we perform aimless shooting simulations, exploring configurations close to the separatrix. We select 500 configurations from umbrella sampling simulations, and 1000 configurations from transition path sampling -- 500 accepted, 500 rejected -- for which we compute the committor. The evaluation of the committor is performed by repeated, independent simulations for which initial velocities are drawn from the Maxwell-Boltzmann distribution, and the system is evolved in the $nVT$ ensemble, until a basin is reached. For each configuration, we perform 200 simulations. In the end, we obtain a dataset of 1500 configurations and corresponding committors. 

We are also interested in seeing how well $\Sigma N$ and $\Sigma C$ capture the true transition state ensemble. For that purpose, from the umbrella sampling dataset, we select configurations at the putative transition state for both simple collective variables using the following conditions: $\Sigma N = 8$ (85 configurations), $7 \leq \Sigma N \leq 9$ (227 configurations), $8 \leq \Sigma C \leq 10$ (153 configurations), and $7 \leq \Sigma C \leq 11$ (259 configurations). These conditions are selected according to the free energy profiles, and correspond to configurations at or close to the transition state ensembles according to both collective variables. Results, in the form of distributions of committor values for these sets of configurations, are reported in Fig. \ref{fig4}(c,d). A good CV would lead to a sharp, unimodal distribution centered at $p(B|\mathbf{X})=0.5$, meaning that its putative transition state ensemble overlaps well with the actual one. Here, the distributions largely lean towards the dissociated state: that most configurations supposed to be transition states according to these CVs are in fact close to one of the metastable states. This means that $\Sigma N$ and $\Sigma C$ lack important information mandatory to appropriately describe the transition between both states. 

From our configuration-committor dataset, we optimize various KRR models, based on different CV subspaces: $\Sigma N$, $\Sigma C$, the pairwise interaction energy between all solute particles ($\Sigma V_{11}$), a combination of $\Sigma C$ and individual solute coordination numbers for all solute particles ($\mathbf{C}$, $d=21$), and the permutation invariant vector\cite{gallet2013} of all solute particles (PIV, $d=190$). The PIV consists of all pairwise distances of the group of atoms, sorted to enforce permutation invariance. While a cutoff function is usually used on the distances to allow focusing on a certain range, here we simply use the inverse distance, which smoothly converges to zero with increasing distance. The detailed procedure for model optimization is reported in Appendix \ref{a:opt}. 

\begin{figure}[h!]
\centering
\includegraphics[width=\textwidth]{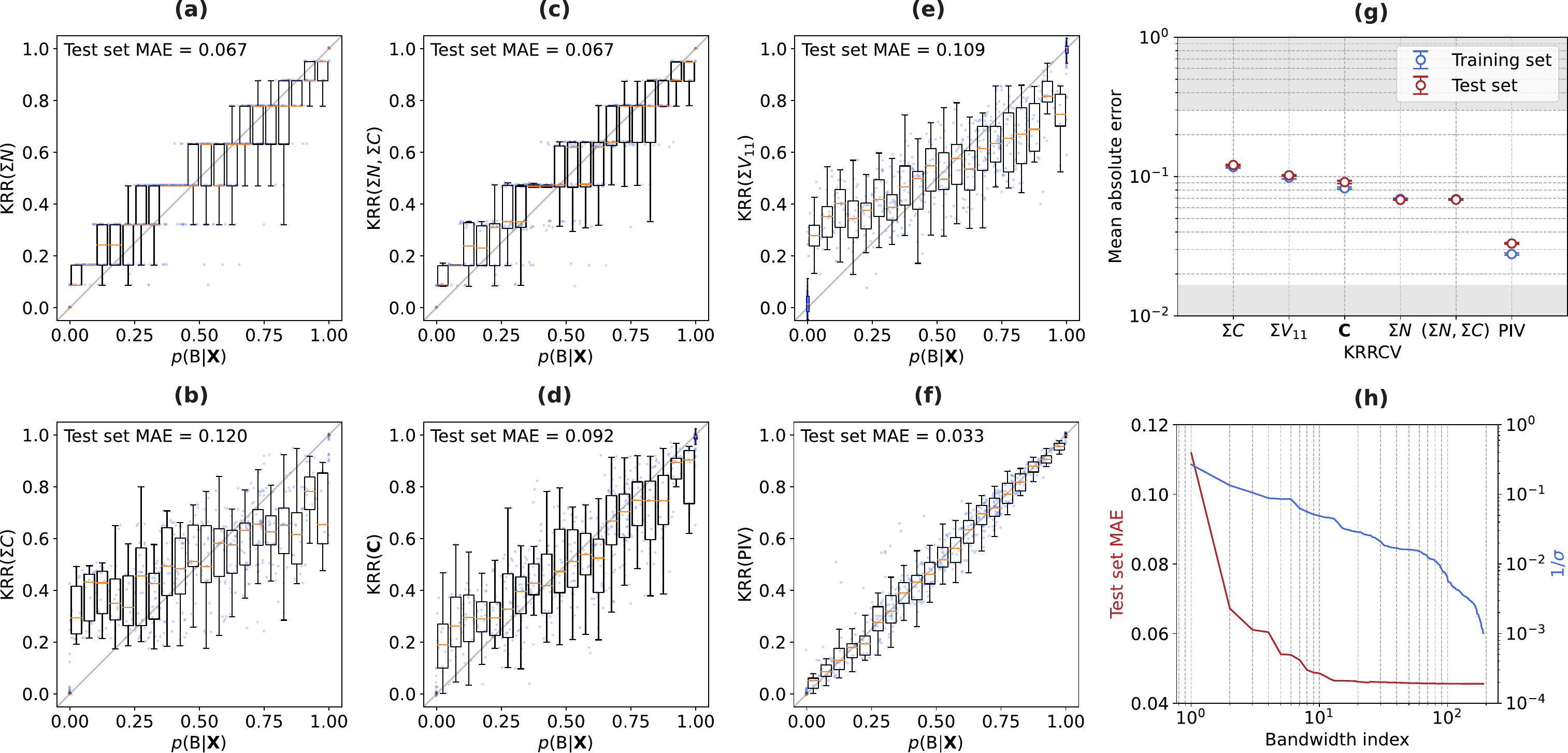}
\caption{KRRCV models for the Lennard-Jones precipitation example. (a-f) Performance on the test set of models based on various collective variables. Small blue circles represent individual data points. The data is also shown using boxplots (the orange line is the median, the box extends from the first to the third quartile of the distribution, whiskers extend from the box edges to the last data point smaller or larger than $\pm 1.5$ times the interquartile range). Data is partitioned in 0.05 wide bins, with two additional blue bins for the basins. (g) Training and test set mean absolute error, averaged over ten dataset splits, plotted as a function of the selected collective variable. Gray boxes show bounds as described in the main text. (h) Sorted inverse bandwidths (blue line), and corresponding test set mean absolute error of reduced models (red line) as a function of the number of included components. }
\label{fig5}
\end{figure}

Performances of the various KRR models are reported in Fig. \ref{fig5}(a-f), comparing test set committor values with model predictions. Training and test set mean absolute errors are reported in Fig. \ref{fig5}(g). Two bounds on the MAE values are derived, and represented on Fig. \ref{fig5}(g) as gray boxes. The upper bound corresponds to the MAE of a na\"{i}ve model always returning the average of its training set. The lower bound, which represents a limit on the meaningful accuracy a model can reach, is due to the uncertainty on the committor estimation, due to the finite number of trials. Here, as demonstrated in Appendix \ref{a:bd}, the MAE on the test set distribution of committor values calculated with 200 trials is about 0.017. First, from Fig. \ref{fig5}(g), it is apparent that CVs rank differently in correlating with the committor. The worst performing ones are $\Sigma C$ and $\Sigma V_{11}$; looking at Fig. \ref{fig5}(b,e), they seem weakly positively correlated with the committor, with larger deviations closer to the basins ($p(B|\mathbf{X})=0$ and 1). $\Sigma N$ (Fig. \ref{fig5}(a)) performs better, in particular close to the basins. Combining $\Sigma N$ and $\Sigma C$ (Fig. \ref{fig5}(c)) as a two-dimensional CV does not seem to improve the performances, which means that all the relevant information relative the the transition in $\Sigma C$ is already captured by $\Sigma N$. The more complicated list of solute coordination numbers $\mathbf{C}$ (Fig. \ref{fig5}(d)) performs as well as $\Sigma N$. Contrarily to the latter, it is however a continuous quantity, which makes it appropriate for biaising. Finally, the PIV based KRR model (Fig. \ref{fig5}(f)) performs significantly better than all other collective variables, with a test set MAE close to the lower bound. A key difference between $\mathbf{C}$ and the PIV collective variables, is that they are composed respectively of single-particle (an atom's coordination number) and two-particle (an interatomic distance) descriptors. While this leads to having to handle a descriptor of much larger dimensionality, the difference in correlation to the committor hints to the importance of two-particle features in properly describing a model precipitation transformation. 

An analysis of the bandwidth components of the PIV-based model (Fig. \ref{fig5}(h)) reveals that most of the relevant information is contained in about 10 pairwise distances, out of 190 in total. The analysis proceeds in the following way: inverse bandwidths are sorted in decreasing order; the largest valued components are the most important. Then, models containing only the $n$ ($1 \leq n \leq 190$) most important components are trained. Finally, the test set MAE of each model is plotted as a function of the number of components included. There is a good correlation between inverse bandwidth values and the performance of the resulting reduced model. This shows that the descriptor can be significantly compressed, allowing to bypass the costly evaluation of all pairwise distances. 

\begin{figure}[h!]
\centering
\includegraphics[width=\textwidth]{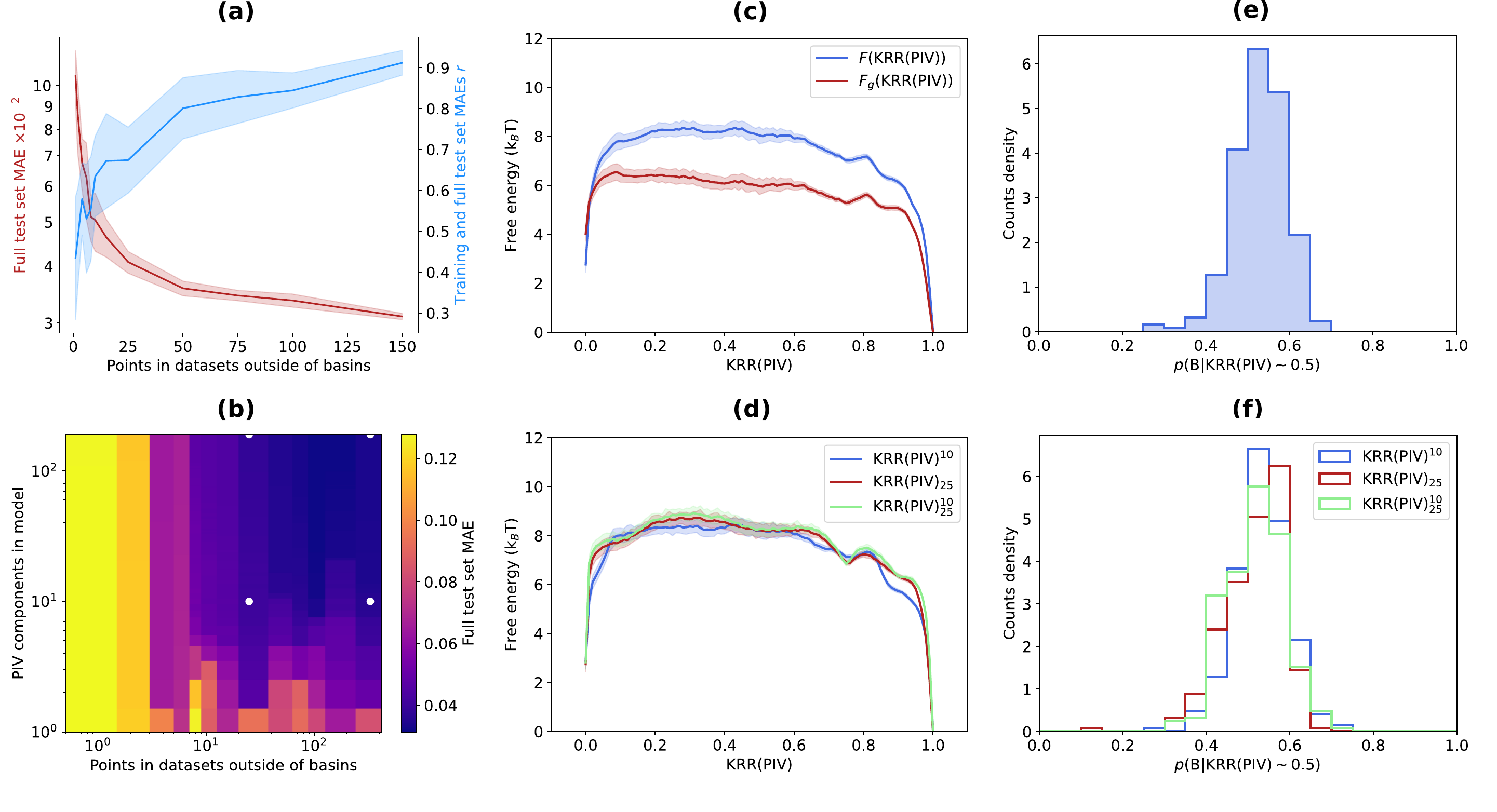}
\caption{Performance of various KRR(PIV) models. (a) Evolution of the full test set MAE as a function of the number of points in the reference and training sets not in metastable basins (red curve). Also shown is the corresponding evolution of the Pearson correlation coefficient betwen the training set and full test set MAEs (blue curve). Results are averaged for 20 dataset splits. (b) Full test set MAE as a function of the number of PIV components included in the model (sorted by bandwidth importance), and number of points in datasets outside of basins. White dots correspond to selected models for further analyses. (c) Free energy and geometric free energy profiles along KRR(PIV). (d) Free energy profiles along reduced KRR(PIV) models. (e) Distribution of committor values for configurations at the putative transition state ensemble of KRR(PIV). (f) Same as (e), for reduced KRR(PIV) models. }
\label{fig6}
\end{figure}

In addition, it is beneficial to minimize the amount of reference configurations in the kernel ridge regression model, most importantly to reduce the amount of committor evaluations, but also to  decrease training and inference times. To this end, we trained PIV models with varying amounts of data points outside of basins (both for reference and training sets), keeping the number of points in basins constant, where the committor can be evaluated at no cost. Results are presented in Fig. \ref{fig6}(a,b). As expected, predictions are poor for datasets only including basin configurations. When increasing the amount of points outside of basins, the accuracy of the model rapidly increases. In addition, the correlation between the reduced training set and full test set MAEs steadily increases. From Fig. \ref{fig6}(b), it is clear that a range of models with reduced dimensionality or data points can achieve similar performance on the full test set. Eventually, we selected four models, represented as white points on Fig. \ref{fig6}(b), for which we performed subsequent tests: the original one (190 PIV components and 330 data points), one with only 10 PIV components preserved (KRR(PIV)$^{10}$), one with only 25 data points outside of basins (KRR(PIV)$_{25}$), and one fully reduced model (KRR(PIV)$^{10}_{25}$). Ideally, selecting an appropriate number of reference points would not require obtaining a large test set. We therefore propose in Appendix \ref{a:modsel} an empirical approach for model selection based on training set error and noise level. Finally, modifying the loss function for bandwidth optimization by increasing the weight on data points outside of basins could allow to obtain satisfying models with even less reference configurations. 

To further confirm the quality of the newly obtained CVs, we perform committor analyses. To sample new configurations at the putative transition state ensembles of KRRCVs, we implement the approach in the \texttt{hack-the-tree} branch of PLUMED\cite{tribello2014,plumedhtt}. Enhanced sampling simulations are automatically enabled as long as the input collective variables forming the subspace are differentiable; we therefore performed umbrella sampling on all four KRRCV models based on the PIV, and obtained datasets constrained at $0.45 \leq \xi \leq 0.55$. We then compute the committor of the sampled configurations, for both collective variables; committor distributions are reported in Fig. \ref{fig6}(e,f). These are in stark contrast with those presented in Fig. \ref{fig4}(c,d). For all four models, the distribution is unimodal and sharply peaked at 0.5, which demonstrates that the projection of the configuration space onto this collective variable preserves the features of the transition state ensemble. Furthermore, we evaluate free energy profiles along these CV models using the previously obtained unbiased molecular dynamics simulations; results are shown, for the original PIV model, in Fig. \ref{fig6}(c). The profile, mostly flat, and rapidly varying close to the basins, is typical of committor approximations\cite{mullen2014}. Profiles obtained for the reduced models (Fig. \ref{fig6}(d)) are very similar. The ensemble-averaged gradients along Cartesian coordinates of $\Sigma N$, $\Sigma C$, and KRR(PIV) are reported in Fig. \ref{a:lj:fig0}. These gradients are used to compute the geometric free energy profile, and intuitively, should be zero in metastable basins, and for monotonous CVs like the committor, maximal at the transition state ensemble. Obviously, this is here only achieved by KRR(CV). Finally, the free energy of activation values for various collective variables (reported in Table \ref{t:deltaf}) are significantly different, even when enforcing gauge-invariance in the free energy profile definition. Additional information is therefore encoded in the KRR(PIV) model, for which $\Delta F^{\ddag}$ is maximized. Since the reaction rate is independent of the collective variable, using the Eyring-Polanyi equation, the transmission coefficients of $\Sigma N$ and $\Sigma C$ should be smaller than the one of KRR(PIV). Recrossings due to a poor choice of collective variable (and therefore separatrix) are therefore reduced. 

\begin{table}[h!]\label{t:deltaf}
\caption{Free energy of activation, for both free energy profile definitions, and for three collective variables. The error estimates are 95\% confidence intervals computed through uncertainty propagation, assuming zero covariance. }
\centering
\begin{tabular}{c c c}
  \hline\hline
  $\xi$ & $\Delta F^{\ddag}$ ($k_B T$) & $\Delta F_g^{\ddag}$ ($k_B T$) \\
  \hline
  $\Sigma N$ & $1.86 \pm 0.21 $ & $1.36 \pm 0.21$ \\
  $\Sigma C$ & $2.61 \pm 0.21 $ & $1.48 \pm 0.22$ \\
  KRR(PIV)   & $5.59 \pm 0.21 $ & $2.52 \pm 0.23$ \\
  \hline\hline
\end{tabular}
\end{table}

\begin{figure}[h!]
\centering
\includegraphics[width=\textwidth]{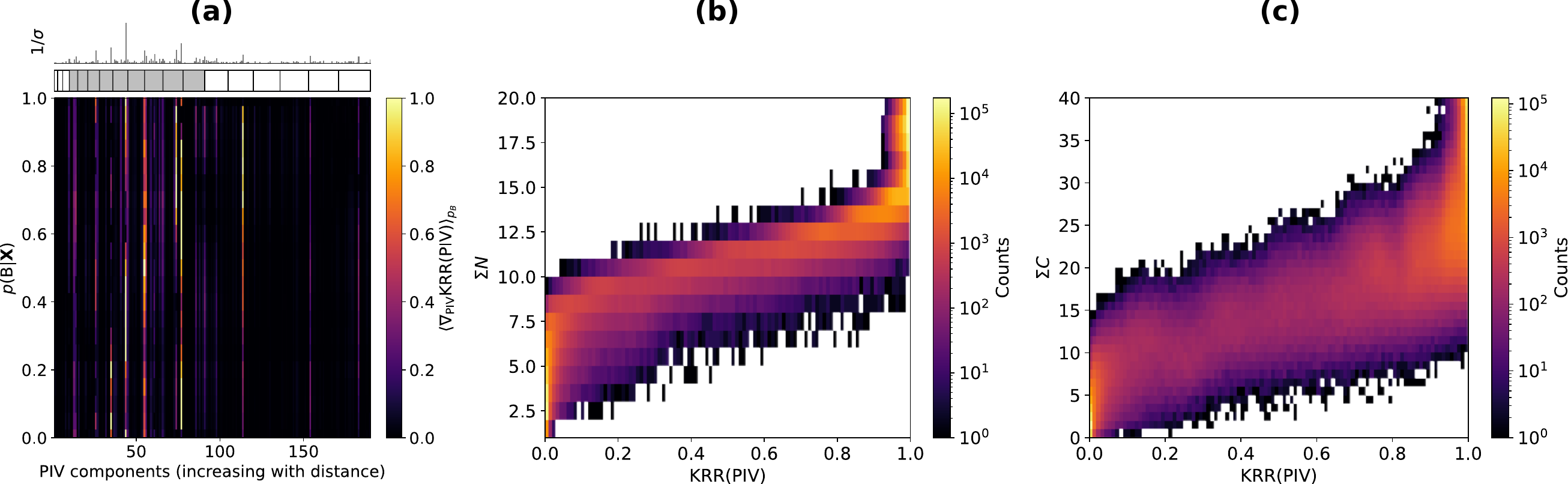}
\caption{(a) Absolute gradient in PIV space of the KRR(PIV) model, averaged and normalized along $p(\text{B}|\mathbf{X})$ bins computed over the test set. The greater the PIV component, the greater the distance out of all pair distances of solute particles. Upper panels: inverse bandwidth values for each PIV components, and corresponding number of particles (white and gray boxes), ranging from 2 (left) to 20 (right) particles. The gray area, from 5 to 14, corresponds to the range between basins definition. (b,c) 2d histograms of configurations sampled from unbiased molecular dynamics, along KRR(PIV) and either $\Sigma N$ (b) or $\Sigma C$ (c). }
\label{fig7}
\end{figure}

A kernel ridge regression approach also ensures the interpretability of the model, since a single parameter (the bandwidth) rates the importance of each component of the collective variable space. For more involved estimators, such as neural networks, this is far from being the case. Obviously, in the end, interpretability will be dictated by the abstractness of the selected collective variable: it is easier to extract knowledge from a small list of simple, intuitive descriptors than from broad, general representations such as a PIV. In the top panel of Fig. \ref{fig7}(a), we display the inverse bandwidth values of the KRR(PIV) model, along with the corresponding number of particles in the PIV displayed as white or gray boxes (the number of distance components grows as $N(N-1)/2$, where $N$ is the number of particles in the system). Most of the important bandwidths are in a range corresponding to 5 to 14 particles in the largest cluster (gray boxes), which is precisely the condition on $\Sigma N$ used for basin definition. In addition, we also evaluate the gradient in PIV space of the KRR(PIV) model along $p(\text{B}|\mathbf{X})$, by binning the test set configurations. The absolute gradient, normalized for each committor bin, is displayed in the bottom panel of Fig. \ref{fig7}(a). This analysis allows to identify the most important PIV components along the reaction process. Although there is some dispersion, along the transition from the dissociated state to the associated state, important components seem to concern larger distances with increasing transformation progress. 

Finally, we can use the KRR(PIV) as a proxy to the committor, to analyze configurations along the reaction process. To this end, we perform an analysis over configurations sampled from unbiased molecular dynamics, here mostly to assess the quality of simple collective variables ($\Sigma N$ and $\Sigma C$) in describing the transformation. We emphasize that a much more thorough analysis using other collective variables can be performed easily, if the interest is in understanding the transformation process, since an inexpensive-to-compute proxy to the committor is now available. 2d histograms along KRR(PIV) and either $\Sigma N$ (b) or $\Sigma C$ (c) are presented in Fig. \ref{fig7}(b,c), computed over the $6 \cdot 10^6$ configurations sampled. Looking at the distribution of $\Sigma N$ at KRR(PIV)$=0.5$, it is interesting to notice that for most configurations, $\Sigma N \approx 10 - 11$, while the top of the barrier shown in Fig. \ref{fig4}(a) corresponds to $\Sigma N = 8$. This also explains the committor distribution in Fig. \ref{fig4}(c), skewed towards the dissociated state. Overall, for constant KRR(PIV) values, the $\Sigma N$ distributions are rather unimodal, although skewed towards large values. For $\Sigma C$, the distributions are very broad over the whole transformation. The basins also overlap with the transition zone. This translates to poor correlation with the committor, and inappropriateness in describing the transformation under scrutiny. 

\section{Ion association in solution: LiF in water}\label{s:lif}

Finally, we investigate the association of Li$^+$ and F$^-$ ions in solution; this mechanism is an important precursor of the formation of the solid electrolyte interphase (SEI) in Li-ion batteries~\cite{peled2017,alzate2022}, its understanding is therefore crucial to design devices with greater lifetimes. We select water as a solvent, both for simplicity, and because an LiF-rich SEI forms at the negative electrode in Li-ion batteries based on aqueous electrolytes~\cite{droguet2021}. LiF, being composed of the smallest monovalent inorganic ions, has a very high lattice energy, and is therefore only very moderately soluble in water (about 1 g/L at room temperature~\cite{jones2009}). 
From the fundamental point of view, the association of atomic, monovalent ions in water is a well-known illustration of the counter-intuitive failure of simple collective variables -- namely the interionic distance -- in describing transformations~\cite{geissler1999,ballard2012,mullen2014}. Arguments put forward include solvent degrees of freedom correlated with the transition as well as inertial effects, \textit{i.e.} when the time-derivative of degrees of freedom has to be accounted for. 

We follow a similar sampling strategy as in section \ref{s:lj}: we first converge the free energy profile along the interionic distance ($r$) by running unbiased molecular dynamics simulations. Then, we sample configurations with $r=r^* \approx 2.63$ \AA, the putative transition state along this collective variable, using umbrella sampling. Finally, we sample configurations near the true separatrix using aimless shooting. Committors are then computed using repeated molecular dynamics for configurations sampled using both methods. Details regarding all computational steps are included in Appendix \ref{a:lif}. The ions are modeled using the Joung-Cheatham\cite{joung2008} potential, and water using the SPC/E rigid model\cite{berendsen1987}. 

\begin{figure}[h!]
\centering
\includegraphics[width=\textwidth]{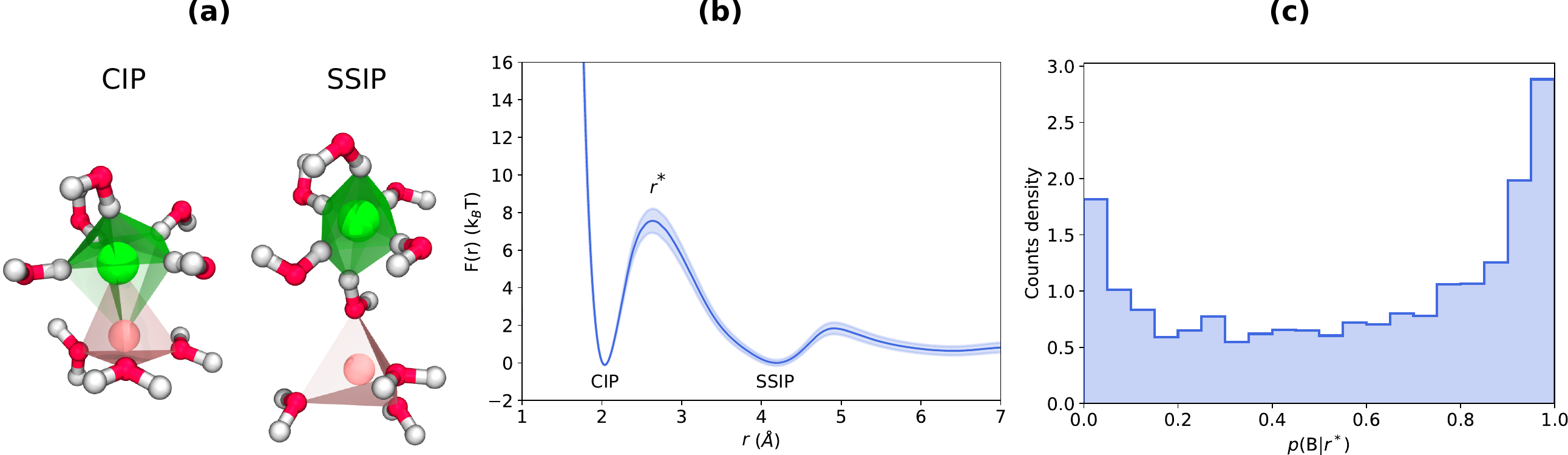}
\caption{LiF association in water. (a)  Contact ion pair (CIP) and solvent-separated ion pair (SSIP). Pink, green, red, and white spheres respectively represent Li, F, O, and H atoms. (b) Free energy profile along the interionic distance ($r$). (c) Distribution of committor values for configurations at the putative transition state ensemble of $r$ sampled using umbrella sampling. }
\label{fig8}
\end{figure}

The free energy along the interionic distance is reported in Fig. \ref{fig8}(b). At short distance, ions form a first metastable state, the contact ion pair (CIP). Separated from the CIP by a free energy barrier of more than $7 k_B T$ is the solvent-separated ion pair (SSIP) metastable state, where one water molecule coordinates both ions. Further increasing the distance leads to completely separating the ions. We define both states using the following criteria: CIP, state A: $r \leq 2.036$ \AA, SSIP, state B: $r \geq 4.190$ \AA. These distances correspond to the bottom of the free energy wells for both states. Other state definitions are possible, such as the interionic distances corresponding to the minima of the wells plus $k_B T$, but a stringent state definition allows to clearly separate the transition regime and the metastable states. The distribution of committor values for configurations sampled at $r^*$ using umbrella sampling is reported in Fig. \ref{fig8}(c). The distribution is bimodal, with sharp peaks at $p(\text{B}|r^*) = 0$ and 1, and mostly flat otherwise. Similar distributions have been recovered for other monovalent ions in water, using different interatomic potentials\cite{geissler1999,ballard2012,mullen2014}. 

\begin{figure}[h!]
\centering
\includegraphics[width=\textwidth]{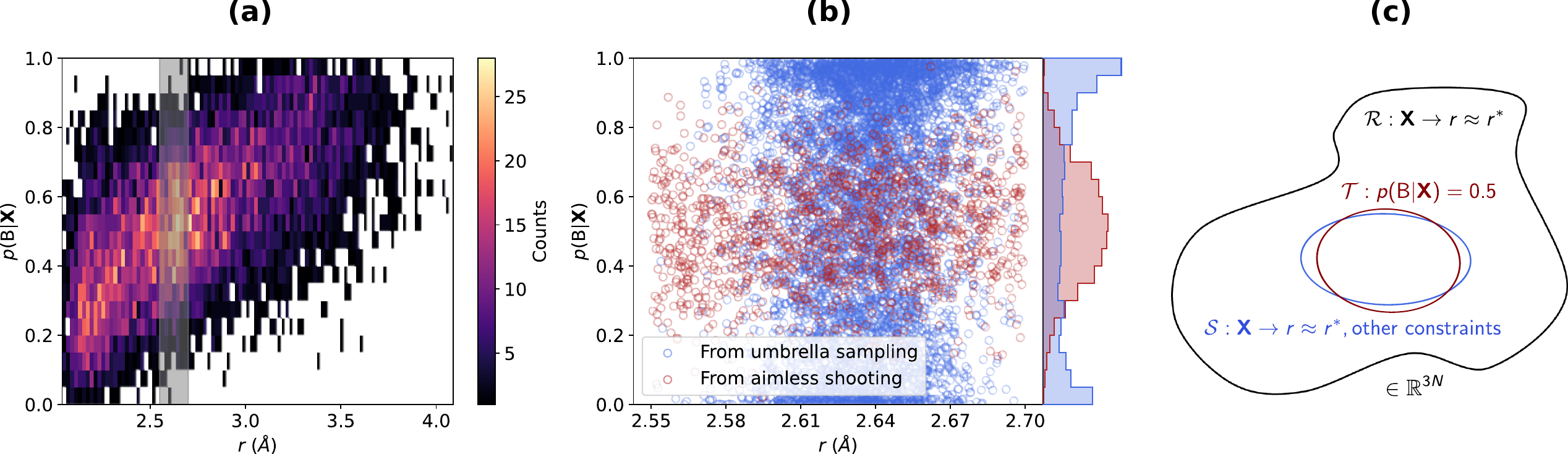}
\caption{(a) Correlation between $r$ and $p(\text{B}|\mathbf{X})$ for configurations sampled from aimless shooting simulations. The gray zone corresponds to $r \in [2.55, 2.70] \approx r^*$. (b) Correlation between $r \approx r^*$ and $p(\text{B}|\mathbf{X})$ for configurations sampled from aimless shooting (red circles) and umbrella sampling (blue circles) simulations. Distributions are also displayed on the right side, with matching colors. (c) Hypothesis regarding the hierarchy of ensembles in configuration space.}
\label{fig9}
\end{figure}

Interestingly, configurations sampled close to the true transition state ensemble correlate fairly well with the interionic distance. This is demonstrated in Fig. \ref{fig9}(a), showing results from aimless shooting simulations. At $r^*$, configurations sampled from aimless shooting are distributed in a unimodal fashion centered at $p(\text{B}|\mathbf{X}) = 0.5$, in stark contrast with configurations extracted from umbrella sampling, as shown in Fig. \ref{fig9}(b). We also verify that the umbrella sampling distribution corresponds to the the distribution under the same conditions obtained from unbiased equilibrium simulations, in Appendix \ref{a:pbdumd}. This could explain why collective variable optimization techniques based exclusively on transition path sampling data show fairly good correlation between the committor and the interionic distance\cite{jung2023} -- not forgetting that the ions and interatomic potentials are not the same. From these observations, we formulate the following hypothesis regarding what is lacking for the proper description of the dissociation process, schematized in Fig. \ref{fig9}(c): the subset of configuration space $\mathcal{R}$ matching the $r \approx r^*$ criterion includes the transition state ensemble $\mathcal{T}$. A subset $\mathcal{S} \subset \mathcal{R}$ approximating $\mathcal{T}$ would match the $r \approx r^*$ condition, with additional unknown constraints on degrees of freedom related to water. Identifying these constraints has been the purpose of many studies\cite{geissler1999,ballard2012,mullen2014,jung2023}. 

\begin{figure}[h!]
\centering
\includegraphics[width=\textwidth]{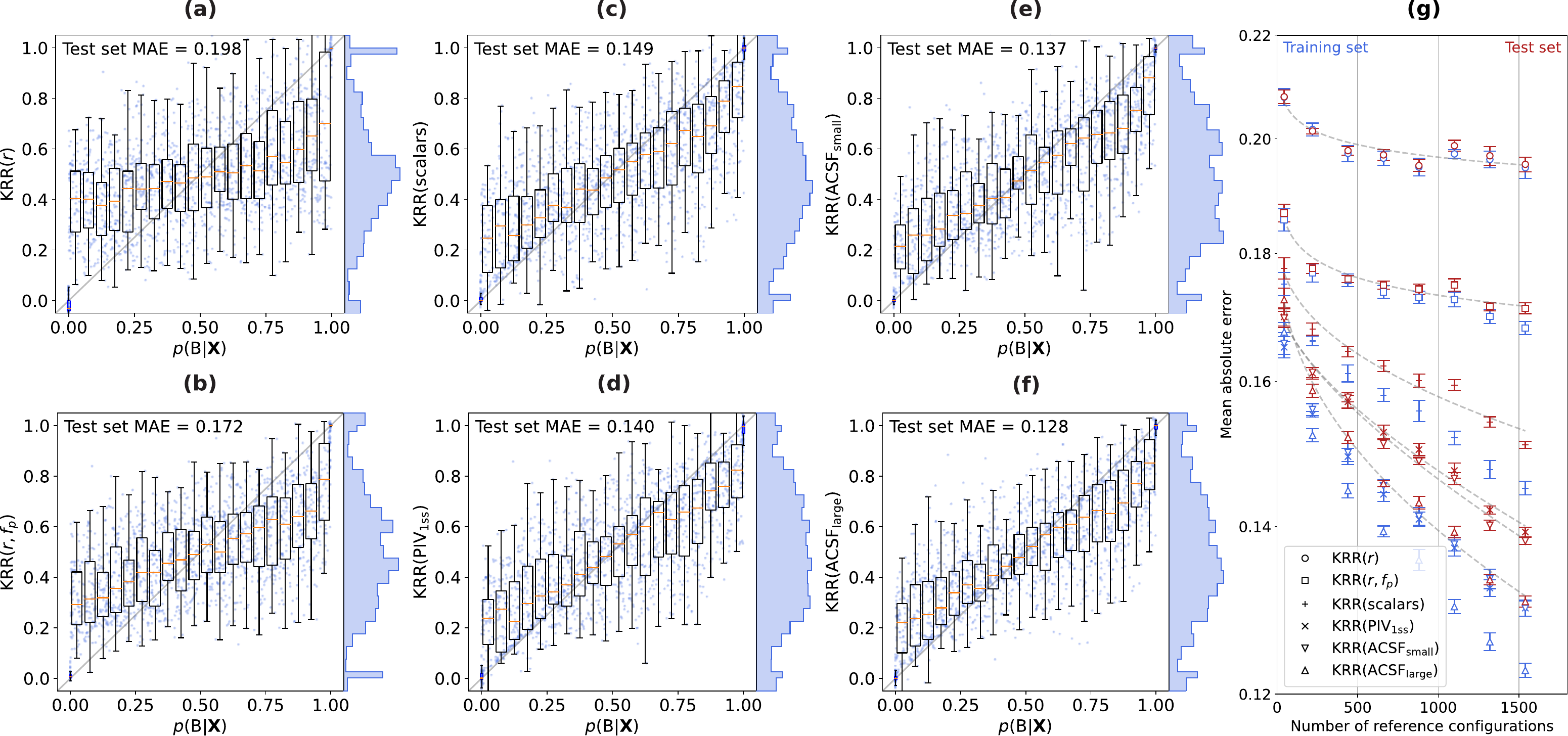}
\caption{KRRCV models for LiF association in water. (a-f) Performance on the test set of models based on varius collective variables. Small blue circles represent individual data points. The data is also shown using boxplots; data is partitioned in 0.05 wide bins, with two additional blue bins for the basins. These are the bins used to generate homogeneous datasets. Distributions of estimated committor values are also provided: an optimal estimator would return a constant distribution, since the test set is homogeneous. (g) Training and test set mean absolute errors, averaged over ten dataset splits, plotted as a function of the selected collective variable. Gray dashed lines are power law fits on the test set MAEs. }
\label{fig10}
\end{figure}

Following these observations, we train six KRRCV models based on different collective variables: i) $r$ only ($d=1$), ii) $r$ and $f_p$, the interionic force projected on the vector connecting both ions ($d=2$), iii) a list of scalar collective variables ($r$, Li$^+$ and F$^-$'s hydrogen and oxygen coordination numbers, the number of water molecules coordinating both ions, and the solvent-contributed Madelung potential\cite{kattirtzi2017} on Li$^+$ and F$^-$ ($d=8$), iv) the PIV of the subsystem composed of both ions and their first coordination sphere, the four closest oxygens to Li$^+$ and the six closest hydrogens to F$^-$ ($d=66$), v) a compact set of atom-centered symmetry functions\cite{behler2007} centered on both ions (ACSFs) designed for aqueous systems\cite{schran2021} ($d=90$), and vi) a larger set of ACSFs automatically designed for organic matter\cite{bircher2021,imbalzano2018} ($d=595$). Reference, training and test sets are homogeneously distributed  in terms of committor values (using 22 bins, with half of the configurations from umbrella sampling, and the other half from aimless shooting). We include basin configurations sampled from equilibrium simulations. The models are trained with a variable number of reference configurations, from 44 to 1540 (2 to 70 per bin). Training and test sets have a fixed number of configurations (1760, or 80 per bin). Results are presented in Fig. \ref{fig10}. 

The first thing to notice is that contrary to the precipitation case, here no single model achieves satisfying performance -- although going from $r$ to more involved collective variables certainly leads to significant improvements. Unsurprisingly, the interionic distance alone leads to a model with high variance along the whole range of committor values. The performances near the basins are particularly poor. Adding $f_p$ slightly improves the situation; in particular, the variance is reduced for configurations close to the transition state. The list of six scalar collective variables, as well as the PIV of the first solvation shell, lead to an improved description of the onset of the reversible transformation mechanism. Estimated committor distributions become noticeably flatter. With the more involved ACSF collective variables, in particular with the high-dimensional ($d=595$) ACSF$_{\text{large}}$, the performance close to the transition state becomes acceptable, although configurations close to the basins remain poorly described. Often the committor is first modeled using a switching function\cite{peters2006,jung2023}, with parameters controlling the shape of the actual function. This allows to obtain a quantity which varies more linearly in the collective variable space, which is therefore easier to fit. Here, this did not improve the situation, with performances of transformed models using a sigmoid function (excluding basins in the training) almost identical to non-transformed models. Kernel ridge regression already allows to learn non-linear patterns, although enforcing these patterns, when known \textit{a priori}, usually makes the learning easier. From Fig. \ref{fig10}(g), it is clear that complex collective variables (i) lead to larger overfitting (\textit{i.e.} the inability of a model to perform satisfyingly far from the training data), and ii) require more reference configurations to converge. Increasing the number of reference configurations beyond the range considered here would certainly improve the performances of the high-dimensional collective variable models. The inability to find an accurate model for the association of LiF in water could be explained by two causes: i) structure beyond the first solvation shell is important, since almost all collective variables considered here focus at most on the ions' first solvation shells (apart from the Madelung potentials), and/or ii) structure alone cannot predict the average evolution of the system, \textit{i.e.} inertial effects are important. In the following, we test both hypotheses. 

\begin{figure}[h!]
\centering
\includegraphics[width=0.5\textwidth]{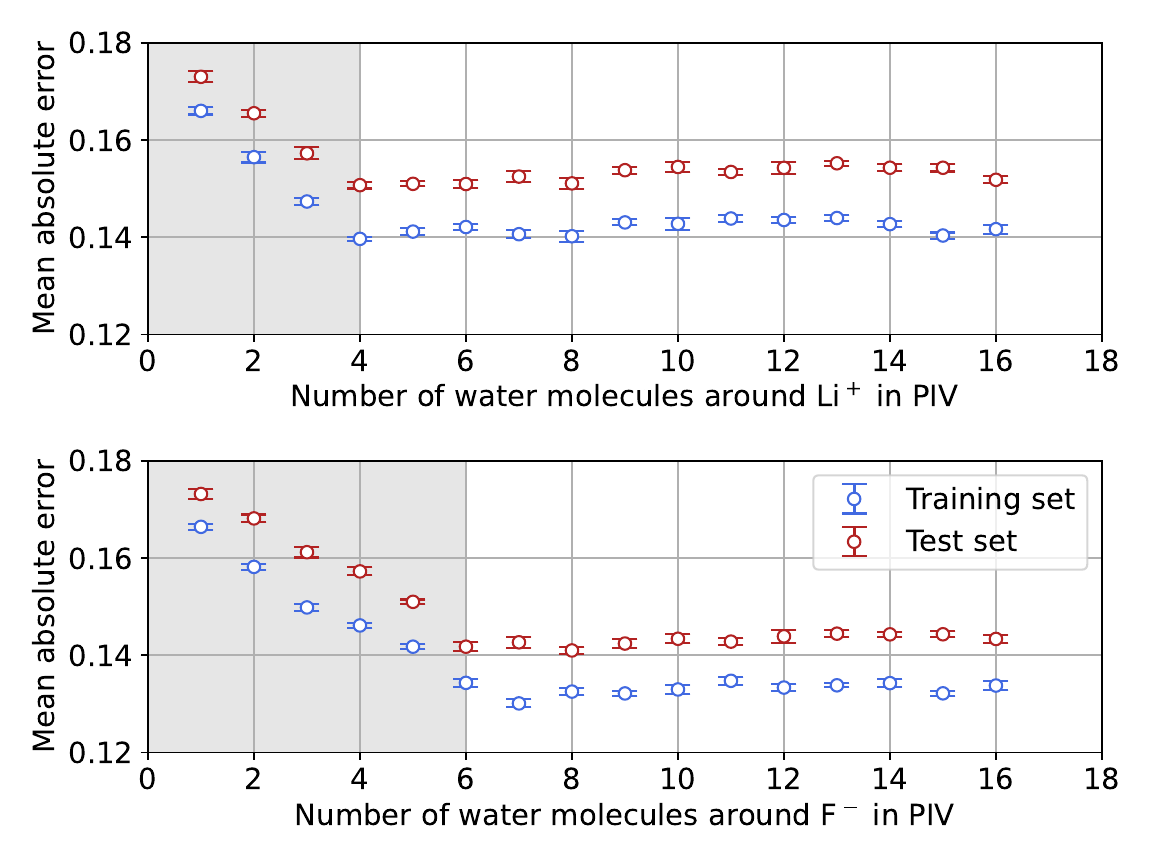}
\caption{Performances of KRRCV models based on the ion-centered PIV with varying number of surrounding water molecules. They gray zones correspond to the first solvation shell of both ions. }
\label{fig11}
\end{figure}

To assess the importance of structure beyond the first solvation shell, we considered PIV-based KRRCV models centered on either the cation or the anion. These always include both ions, and a variable number of water molecules (from 1 to 16), ranked by their distance to the central ion (O-Li$^+$, or H-F$^-$). The dimension of the collective variable ranges from 10 to 1225. Performances of the models, using 1760 reference configurations, are reported in Fig. \ref{fig11}. 

While increasing the number of water molecules initially leads to reducing both training and test set MAEs, which corresponds to including more information correlated to the transformation mechanism, a plateau is rapidly reached. Interestingly, this plateau is exactly reached at a number of water molecules corresponding to the first solvation shell of both ions (4 for the cation, 6 for the anion). Increasing the number of water molecules, and therefore the dimension of the collective variable, does not lead to KRRCV models with degraded performances, as can be seen from the plateau in the MAEs. This means that the optimizer still performs well, for high-dimensional data. Within a PIV representation, there is therefore no information correlated to the LiF association mechanism beyond the first solvation shell. Assuming that the mechanism is similar for NaCl, this finding is both in agreement\cite{wang2022} and disagreement\cite{ballard2012} with previous studies. 

To investigate inertial effects, we start by evaluating their importance in the present setting. Hummer showed\cite{hummer2004} that in the diffusive limit, the committor and the transition path probability $p(\text{TP}|\mathbf{X})$ (the probability that a trajectory passing through configuration $\mathbf{X}$ connects both basins) are related in the following way: 

\begin{equation}
p(\text{TP}|\mathbf{X}) = 2 p(\text{B}|\mathbf{X}) \left( 1 - p(\text{B}|\mathbf{X}) \right).
\end{equation}

\begin{figure}[h!]
\centering
\includegraphics[width=\textwidth]{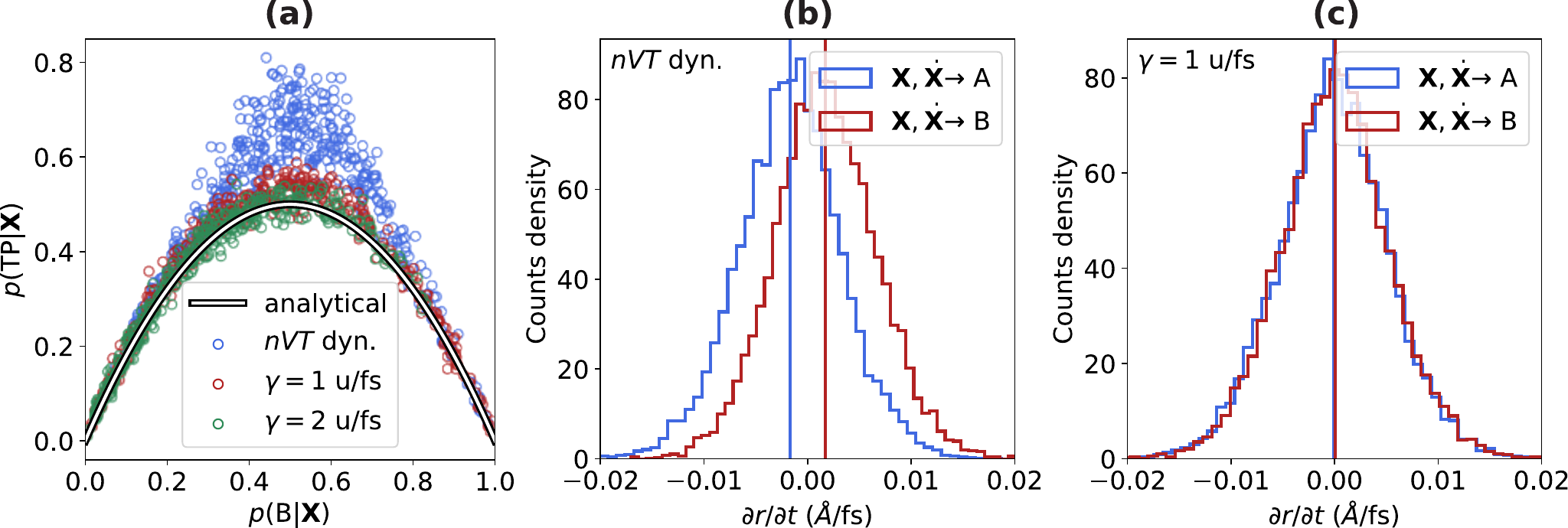}
\caption{Inertial effects: (a) correlation between $p(\text{B}|\mathbf{X})$ and $p(\text{TP}|\mathbf{X})$ for different cases: $nVT$ dynamics using a chain of Nosé-Hoover thermostats (blue circles), two overdamped Langevin dynamics with different friction values (red and green circles), and the limit of infinite friction (black and white line), (b,c) outcome histograms over initial interionic velocity, for $nVT$ dynamics (b), and overdamped Langevin dynamics (c). The red and blue vertical lines correspond to the average of each distribution. }
\label{fig12}
\end{figure}

Away from the diffusive limit, inertial effects are important and lead to a deviation from this analytic result. For 400 configurations sampled from aimless shooting, we compute explicitly $p(\text{TP}|\mathbf{X})$; results are presented in Fig. \ref{fig12}(a) (blue circles, $nVT$ dynamics using a chain of Nosé-Hoover thermostats). A strong deviation from the diffusive limit, maximal at the transition state ensemble, is apparent. This deviation has also been observed for the association of NaCl in water\cite{ballard2012}. Configurations near the transition state ensemble have an enhanced transition path probability. In the diffusive limit, atomic velocities are randomized at each timestep, while under $nVT$ dynamics, the persistence in time of momenta can influence directly the occurrence of a reactive event. To further confirm the importance of inertial effects, we perform 1000 velocity initializations for 20 configurations from umbrella sampling for which $0.475 \leq p(\text{B}|\mathbf{X}) \leq 0.525$, and monitor the initial atomic velocities projected on the interionic distance ($\partial r / \partial t$), as well as the basin the system eventually commits to. Histograms depending on the committment outcome are presented in Fig. \ref{fig12}(b). In the absence of inertial effects, both histograms would overlap, indicating that the outcome is not conditioned on the atomic velocities. However, a clear separation is visible here, with a negative (positive) interionic velocity being more likely to lead to association (dissociation). This is a trivial result if one assumes that the persistence of atomic momenta is non-zero. Considering the reversibility of the equations of motion, and therefore, in particular, of committor trajectories, this result can also be interpreted in a specular way: initial velocities Boltzmann-distributed in each basin lead to reactive trajectories with transition-state velocities having an asymmetric non-Boltzmann distribution. 

\begin{figure}[h!]
\centering
\includegraphics[width=.5\textwidth]{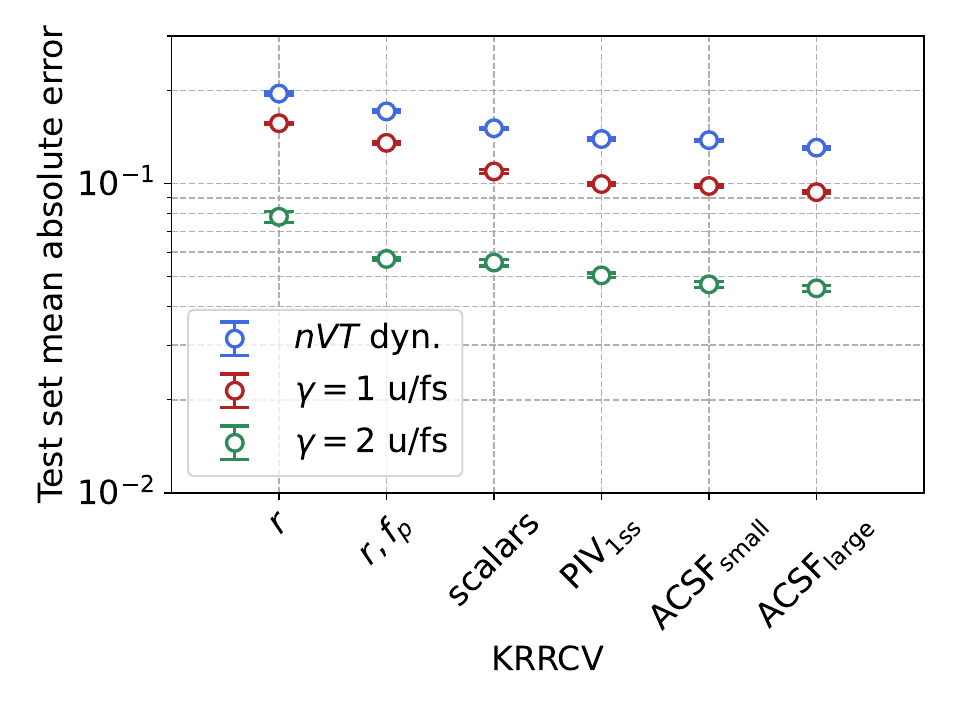}
\caption{Performances of various KRRCV models, for different dynamics. }
\label{fig13}
\end{figure}

Having shown that inertial effects do play a role in the context of LiF association in water, we wish to alter the dynamics in such a way that they are reduced. To this end, we modify the equation of motion of the ions only (\textit{i.e.} that for water is unchanged), following the overdamped Langevin equation: 

\begin{equation}\label{e:ole}
\mathrm{d}\mathbf{X} = - \gamma^{-1} \nabla U \mathrm{d}t + \sqrt{2 k_B T} \gamma^{-1/2} \mathrm{d}\mathbf{W},
\end{equation}

where $\mathbf{W}$ is a $3n$-dimensional white noise term (with $n$ the number of ions), and $\gamma$ is the friction, identical for all ions. We perform simulations with $\gamma = 1 $~u/fs and 2~u/fs, corresponding to inertial timescales -- \textit{i.e.}, timescales for decorrelation of the CV time derivative -- of about 7 fs and 14 fs on Li$^+$, and 19 fs and 38 fs on F$^-$. As demonstrated in Fig. \ref{fig12}(a), increasing the friction parameter leads to a better agreement with the diffusive limit (which corresponds to $\gamma\to\infty$). As shown in Fig. \ref{fig12}(c), the outcome histograms now overlap, signifying that inertial effects are removed. We then evaluate the committor on the umbrella sampling and aimless shooting datasets previously obtained, with both friction values, and train new KRRCV models. Results are presented in Fig. \ref{fig13}. 

The performances of KRRCV models are systematically and significantly improved compared to $nVT$ dynamics. This is valid for each collective variable. Obviously, in the diffusive limit, the interionic distance would be the ideal collective variable, but considering that such an improvement is already obtained at low to reasonable friction values, this demonstrates that inertial effects prevent finding a collective variable only based on the structure that optimally reproduces the committor. In addition, even for $\gamma = 2$~u/fs, the hierarchy of performance between collective variables is preserved. 

There are therefore two effects to take into account when searching for an optimal collective variable for the association of small atomic ions in water: structural degrees of freedom related to the solvent, which contain information correlated with the committor, and inertial effects. 
An approach purely based on the committor in the case where inertial effects are important seems inappropriate, as $p(\text{B}|\mathbf{X})$ does not encode all the necessary information describing the transformation under scrutiny. 

Note that the velocity-dependent committor $p(\text{B}|\mathbf{X}, \dot{\mathbf{X}})$ is a deterministic quantity in the microcanonical ensemble, as it corresponds to a single point in phase space, while it retains a probabilistic nature in presence of a thermostat (see, \textit{e.g.}, Ref.\cite{metzner2006} for a discussion of velocity-dependent committors in transition path theory). 

A possible approach suitably accounting for inertial effects could be based on a committor conditioned on the velocity of a collective variable. This quantity is still a probability, and it would be interesting to determine to which extent inertial effects are important: do they only concern the ions, in which case evaluating $p(\text{B}|\mathbf{X}, \dot{r})$ should be enough, or are the momenta of the water molecules also correlated to the transformation mechanism? To efficiently evaluate $p(\text{B}|\mathbf{X}, \dot{r})$, velocity initialization techniques with fixed values of $\dot{r}$ would have to be implemented. 

\section{Summary and conclusions}\label{s:c}

We have introduced a new method for the construction of optimized collective variables, based on a data-driven generalization of path collective variables. It consists in performing a regression of the committor -- which, in the absence of inertial effects, directly encodes the transformation progress -- starting from a given basis of collective variables, through kernel ridge regression. The final collective variable is one dimensional and differentiable, which makes it appropriate for enhanced sampling simulations. Our approach also allows to rank and compare CVs: the performance of a certain CV subspace choice is evaluated by computing a mean absolute error with respect to a test set of committor values. In addition, using a bandwidth vector in the kernel, with values optimized by minimizing a loss score defined on a training set, allows for interpretability of the resulting model. We apply the method to three different test cases: a two-dimensional toy model, a precipitation model of Lennard-Jones particles, and the LiF association in water. 

From the two-dimensional toy model test case, we learn that a reasonable number of reference configurations is sufficient to obtain an accurate KRRCV model (Kernel Ridge Regression Collective Variable), although how they are sampled is important, with enhanced sampling being a viable strategy for constructing datasets. In addition, the optimization process of kernel bandwidth parameters is shown to be robust even when including many collective variables uncorrelated with the committor. 

For the precipitation model of Lennard-Jones particles, we show that our method allows to obtain an accurate collective variable from a high-dimensional representation of the structure (the PIV). This is promising for further investigation of realistic precipitation, crystallization and nucleation phenomena. The model obtained can be significantly compressed, both in terms of the number of reference configurations and components included, without degrading its performances, evaluated on committor distributions and free energy profiles. In particular, for committor distributions, we were able to perform biased simulations using KRRCVs. The free energy of activation of the best-performing model, extracted from the gauge-invariant free energy profile, is larger than the ones obtained using simple collective variables, hinting at a significantly improved transmission coefficient too. Finally, we show that the model can be interpreted, and is useful to evaluate issues related to less well-performing collective variables. 

For the association of LiF in water, we demonstrate that no information correlated with the transformation mechanism is contained beyond the first solvation shell of the ions. We obtain various KRRCV models performing better than the simple interionic distance. The highest accuracy reached is however moderate, even when resorting to flexible representations such as high-dimensional ACSFs. We show that this is mostly due to inertial effects. In this regime, the committor depends not only on the atomic positions, 
but also on their momenta. 

In situations where inertial effects are important, such as reactions in solution, it therefore seems mandatory to account for the atomic momenta. A solution could be to evaluate committor probabilities conditioned on a collective variable momentum. In this setting, much remains to be understood. For instance, should the momenta be projected onto collective variables in the same way as the structure? Or is there a hierarchy that would allow to use simpler projections for the velocities, compared to the atomic positions; for instance, in the ion association example, could we get away with only conditioning the committor on the interionic distance time derivative? In situations where inertial effects are negligible (precipitation, nucleation, crystallization), the method introduced here allows to help understand transformation mechanisms, select collective variables on solid grounds, and allow for efficient enhanced sampling simulations. 

\begin{acknowledgments}
The authors would like to gratefully acknowledge fruitful discussions with Sara Bonella, Michele Casula, Ludovic Goudenège, Pierre Monmarché, Rocio Semino, Alessandra Serva, Gabriel Stoltz, and Rodolphe Vuilleumier, within the MAterials for Energy through STochastic sampling and high-peRformance cOmputing (MAESTRO) collaboration, hosted at the Institut des Sciences du Calcul et des Données. We also thank Gareth Tribello for his advice in implementing KRRCVs in the \texttt{hack-the-tree} branch of PLUMED. This project received funding from the European Research Council under the European Union's Horizon 2020 research and innovation program (Grant Agreement No.~863473). This work was granted access to the HPC resources of IDRIS under the allocations A0130811069 and A0140901387, and also benefited from the computing resources at Sorbonne Université managed by SACADO. 
\end{acknowledgments}

\appendix
    
\section{Solving the backward Kolmogorov equation using finite elements}\label{a:fe}

The committor $p(\text{B}|(x,y)) $ can be obtained as the solution of the following equation~\cite{Zhang2017}
\begin{equation}\label{e:bke}
\begin{cases}
      \mathcal{L} p(\text{B}|(x,y)) = 0  & (x,y) \in (A\cup B)^c\\
       p(\text{B}|(x,y)) = 0 & (x,y) \in A \\
       p(\text{B}|(x,y)) = 1  &(x,y) \in B 
\end{cases}
\end{equation}
where $\mathcal{L}$ is the infinitesimal generator of the Langevin overdamped dynamics. For the two-dimensional potential $V(x,y)$ of Eq.~\eqref{e:rmb}, it can be expressed as~\cite{Pavliotis2014}
\begin{equation}
    \mathcal{L} f(x,y)= -\frac{\partial V(x,y)}{\partial x} \frac{\partial f(x,y)}{\partial x} -  \frac{\partial V(x,y)}{\partial y} \frac{\partial f(x,y)}{\partial y} +  \frac{1}{\beta} \left(\frac{\partial^2 f(x,y)}{\partial x^2}+2\frac{\partial^2 f(x,y)}{\partial x\partial y}+ \frac{\partial^2 f(x,y)}{\partial y^2}\right)
\end{equation}
for any two-dimensional function $f(x,y)$. Eq.~\eqref{e:bke} can be solved using finite elements methods~\cite{lapelosa2013}. For a finite element basis $\{f_i(x,y)\}_1^N$, the backward Kolmogorov equation can be expressed as a matrix equation,
\begin{equation}
    \label{e:weakform}
    \begin{cases}
        \bm{L} q = 0 & (x,y) \in (A\cup B)^c\\
        q = 0 & (x,y) \in A\\
        q = 1 & (x,y) \in B
    \end{cases}
\end{equation}
where the matrix elements are given by 
\begin{multline}
    \bm{L}_{i,j} = \int_{\mathbb{R}^2} \mathrm{d}x\,\mathrm{d}y\,\left[f_i(x,y) \left( -\frac{\partial V(x,y)}{\partial x} \frac{\partial f_j(x,y)}{\partial x} -  \frac{\partial V(x,y)}{\partial y} \frac{\partial f_j(x,y)}{\partial y} \right)-  \right. \\
    \left. \frac{1}{\beta} \left(\frac{\partial f_i(x,y)}{\partial x}\frac{\partial f_j(x,y)}{\partial x}+\frac{\partial f_i(x,y)}{\partial x}\frac{\partial f_j(x,y)}{\partial y}+\frac{\partial f_i(x,y)}{\partial y}\frac{\partial f_j(x,y)}{\partial x}+ \frac{\partial f_i(x,y)}{\partial y}\frac{\partial f_j(x,y)}{\partial y}\right)\right].
\end{multline}
The committor is then obtained as
\begin{equation}
    p(\text{B}|(x,y))  = \sum_i q_i \, f_i(x,y) 
\end{equation}
The result of Fig.~\ref{fig2}b was obtained using a regular triangular mesh of $79202$ triangles defined on the rectangular domain $[-1.5, 1] \times [-0.5, 2]$ and linear element ($\mathbb{P}1$ Lagrange element). This leads to a finite element basis with $39534$ degrees of freedom once nodes corresponding to region $A$ and $B$ were removed. Computations were performed using the scikit-fem finite element library~\cite{Gustafsson2020}.

\section{KRR model optimization procedure}\label{a:opt}

All optimizations (regarding kernel ridge regression expansion coefficients as well as hyperparameters) are performed using the python/C++ package falkon\cite{meanti2020,meanti2022}. A model python script is included in the Supplemental Material. For a given set of hyperparameters (regularization coefficient and bandwidths), optimal expansion coefficients are obtained by solving Eq. \ref{e:akrr}. Simultaneous optimization of hyperparameters is achieved by iteratively minimizing the MAE on a training set (distinct from the reference set), using the Adam optimizer\cite{kingma2014}, as implemented in PyTorch\cite{paszke2019}, for 100 steps. For a dataset of $N$ entries, the MAE is defined as: 

\begin{equation}
\text{MAE} = \frac{1}{N} \sum_{i=1}^N | p(\text{B}|\mathbf{X}_i) - \text{KRRCV}(\xi_i) |
\end{equation}

\begin{figure}[h!]
\centering
\includegraphics[width=.5\textwidth]{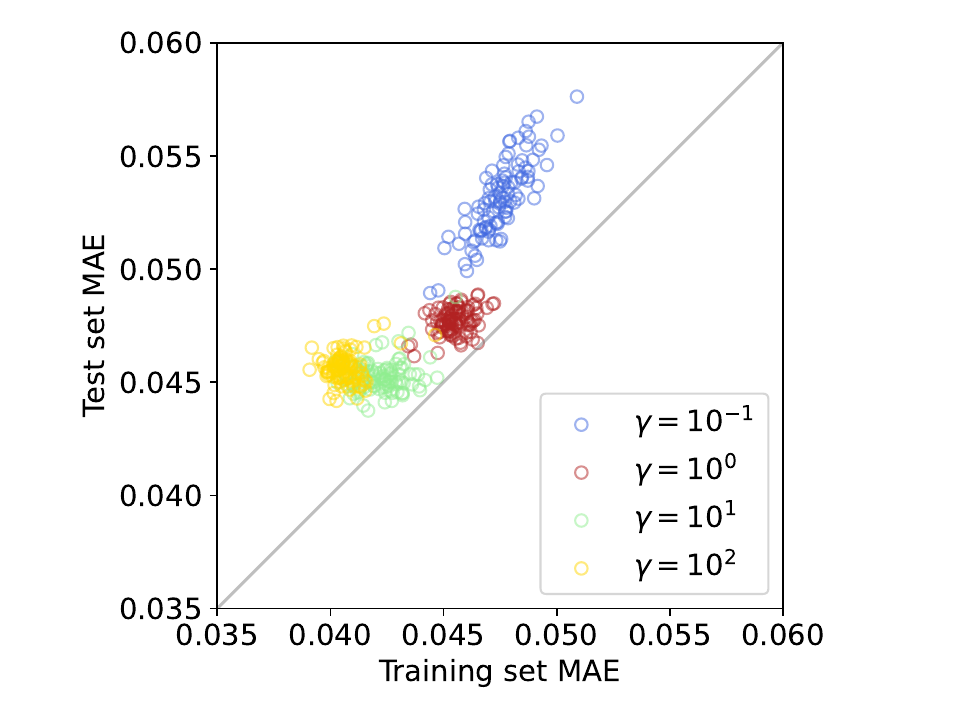}
\caption{Correlation between the training set and test set mean absolute errors, for the Lennard-Jones precipitation case, with the PIV collective variable, for the dataset split 3. Points are colored according to the learning rate $\alpha$; each point corresponds to the final metrics optained for one optimization, starting from randomly selected parameters. }
\label{a:opt:fig1}
\end{figure}

We perform distinct optimizations with different values of the learning rate parameter ($\alpha$): for the rugged Müller-Brown case, $\alpha \in \left[10^{-1}, 10^{0}, 10^{1}\right]$, for the Lennard-Jones case, $\alpha \in \left[10^{-1}, 10^{0}, 10^{1}, 10^{2}\right]$, and for the LiF association case, $\alpha \in \left[10^{-1}, 5 \cdot 10^{-1}, 10^{0}, 5 \cdot 10^{0}, 10^{1}, 5 \cdot 10^{1}, 10^{2}, 5 \cdot 10^{2}\right]$. We partition reference and training sets randomly in 10 different ways, to obtain uncertainty estimates. The test set is always the same. We perform 100 optimizations starting from randomly selected initial hyperparameter values. For each $\gamma$ value, this amounts to 1000 optimizations. Finally, out of all optimized models, we select the one that minimizes the training set error. We observe a positive correlation between the training set and test set error metrics, as shown on Fig. \ref{a:opt:fig1}.

We note that for high-dimensional representations, the optimal models are "ridgeless", meaning that the optimal regularization parameter $\lambda$ approaches zero. This has been discussed before \cite{liang2020} for non-linear kernels, and does not prevent generalization. 

\section{Three wells model potential}\label{a:3w}

\begin{figure}[h!]
\centering
\includegraphics[width=\textwidth]{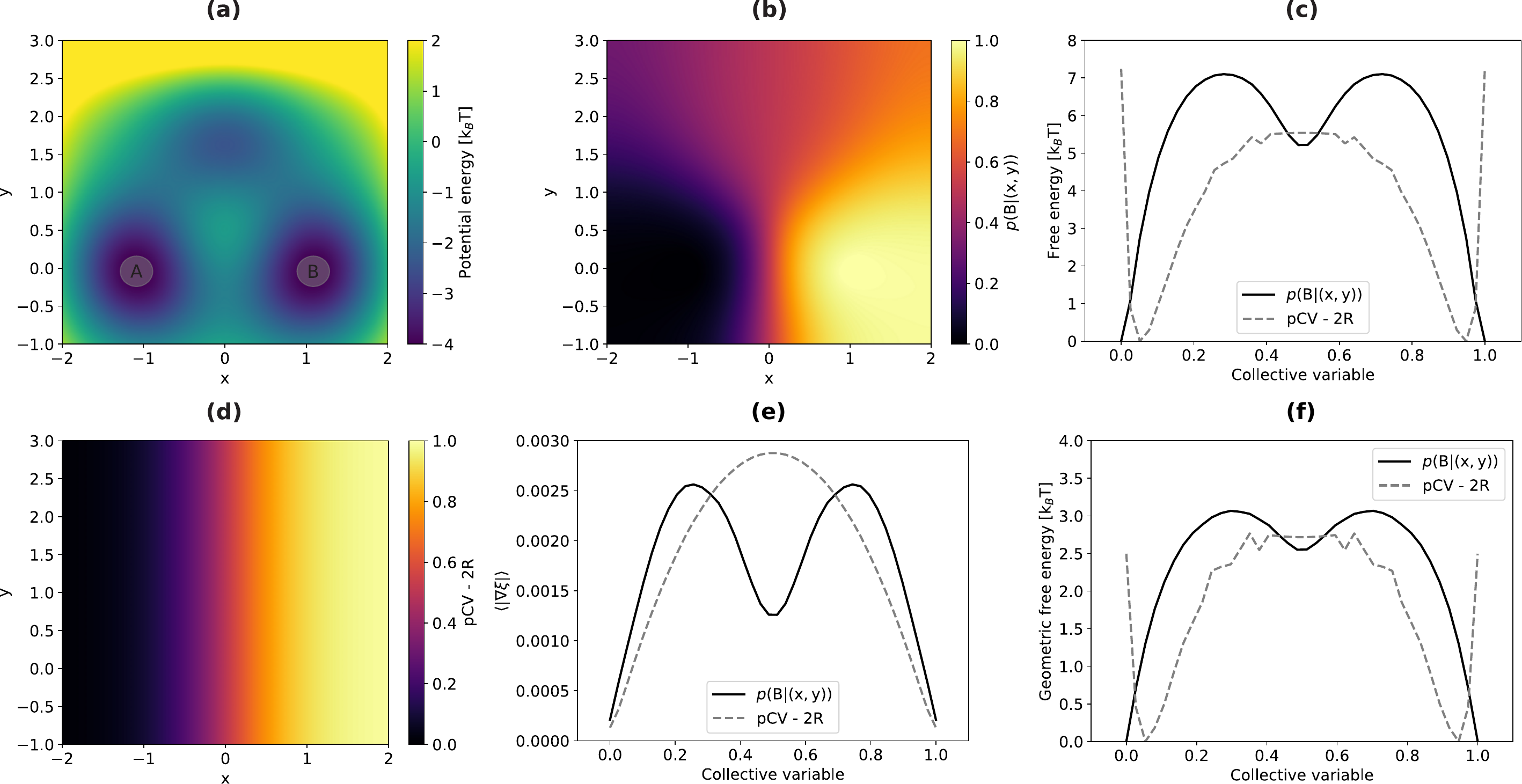}
\caption{(a) The potential energy surface. Shaded circles correspond to the metastable states definition used for the evaluation of the committor, (b) the committor as obtained from the backward Kolmogorov equation, (c) Canonical free energy profiles, (d) the "2R" path collective variable in configuration space, (e) Ensemble-averaged collective variable gradients, (f) Geometric free energy profiles. }
\label{a:3w:fig1}
\end{figure}

In Fig. \ref{a:3w:fig1}, we compare a "2R" path collective variable with the true committor for a two-dimensional potential showing three metastable states. The two deeper ones are labeled as A and B, and the shallowest one is an intermediate state. The free energy profiles differ quite significantly, especially in the vicinity of the intermediate state. 

\section{Müller-Brown potential non-linearly embedded in a five-dimensional space}\label{a:5d}

To complicate the learning process of the committor in Sec. \ref{s:rmb}, we use non-linear transformation to embed the two-dimensional Müller-Brown potential in a five-dimensional space, in a similar way as in Ref. \cite{sun2022}: 

\begin{equation}
V(x_1,x_2,x_3,x_4,x_5) = V(x,y), 
\end{equation}

\begin{equation}
\begin{cases}
x_1 = & x + 0.1 y^2 \\
x_2 = & y - 2x + 3\\
x_3 = & \sqrt{4 |xy|}\\
x_4 = & x^3 - y^2\\
x_5 = & xy^4\\
\end{cases}.
\end{equation}

We train a KRR model using all five dimensions, with 500 reference and training data points. The resulting test set MAE, $\approx 7 \cdot 10^{-3}$, is on par with the one of the model trained on the native two-dimensional representation. 

\section{Precipitation of Lennard-Jones particles: computational details and additional information}\label{a:lj}

\subsection{System generation, initial relaxation}

We begin with an initial configuration composed of 4096 particles arranged on a simple cubic lattice of spacing set to $l=\sigma$, in a cubic box with 16 $\sigma$-long edges. 20 particles, selected randomly, are set to being of type 2 (the larger, precipitating species). The atomic velocities are initialized by drawing from the Maxwell-Boltzmann distribution at $T=1$. A $10^6 \delta t$ simulation in the $npT$ ensemble at $T=1$, $p=1$ is then performed to relax the system and estimate the equilibrium box size. The box size is subsequently fixed at 17.20 $\sigma$. 

\subsection{Unbiased free energy estimates}

Starting from the previously equilibrated geometry, we perform 300 independent simulations in the $nVT$ ensemble with randomized initial velocities, ran for $10^6 \delta t$ for equilibration and $2\cdot 10^7 \delta t$ for sampling, which amounts to a total of $6 \cdot 10^9 \delta t$. Collective variables are computed every $10^3$ time steps. Free energy profiles are then calculated through binning, with error estimates obtained by splitting the whole dataset into six distinct subsets, and computing $95\%$ confidence intervals over the distribution of estimates. We performed identical simulations in the $npT$ ensemble, to verify that box size fluctuations do not significantly influence free energy profiles. 

\subsection{Computing collective variable gradients}\label{a:lj:grad}

We compute the derivatives of the collective variable with respect to the Cartesian coordinates using a second-order central difference scheme, with a displacement of atomic positions set to $0.05 \sigma$: 

\begin{equation}
\frac{\mathrm{d} \xi(x)}{\mathrm{d} x} \approx \frac{\xi(x + 0.05 \sigma) - 2 \xi(x) + \xi(x - 0.05 \sigma)}{(0.05 \sigma)^2}
\end{equation}

\begin{figure}[h!]
\centering
\includegraphics[width=\textwidth]{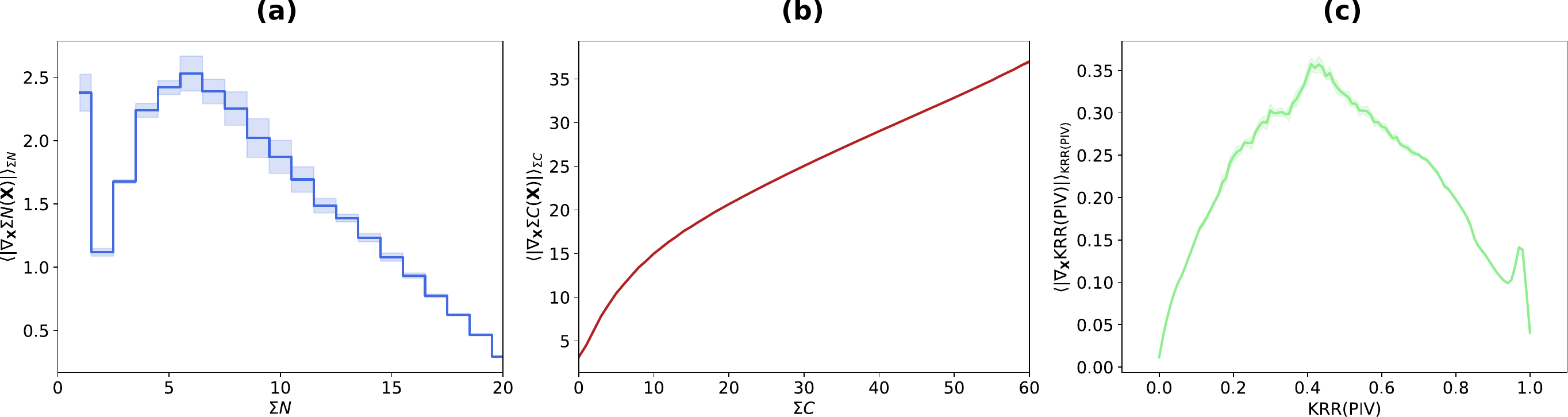}
\caption{Norm of the gradient of collective variables along Cartesian coordinates: (a) $\Sigma N$, (b) $\Sigma C$, (c) KRR(PIV)}
\label{a:lj:fig0}
\end{figure}

\subsection{Umbrella sampling simulations}

To sample configurations biased along $\Sigma C$, we perform simulations with five different harmonic biasing potentials of the form $0.5 k\left( \Sigma C - \Sigma C_0\right)^2$ centered at $\Sigma C_0 = 10, 11, 12, 13, 14$, and with a force constant $k = 10 k_BT$. Trajectories last $10^7 \delta t$ and configurations are sampled every $10^5$ steps; the first configuration is discarded to allow for equilibration. We therefore sample a total of 500 configurations. 

To sample configurations at the putative transition state ensembles of KRR($\mathbf{C}$) and KRR(PIV), we perform simulations with a harmonic biasing potential centered at KRR($\xi$)$_0=0.5$, and with a force constant $k = 10^4 k_BT$. Trajectories last $2.5 \cdot 10^6 \delta t$ and configurations are sampled every $10^4$ steps; the first configuration is discarded to allow for equilibration. We therefore sample a total of 250 configurations for each collective variable. 

\subsection{Transition path sampling: brute force}

\begin{figure}[h!]
\centering
\includegraphics[width=.5\textwidth]{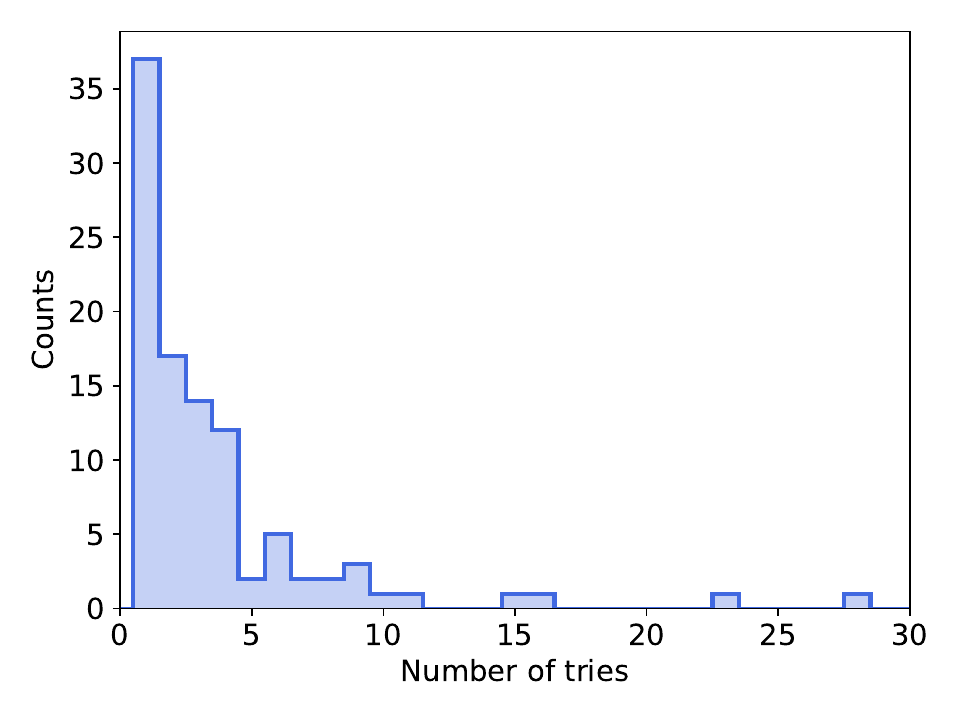}
\caption{Number of tries needed to connect both basins in brute force transition path sampling, for 100 starting configurations sampled from an umbrella sampling simulation.}
\label{a:lj:fig1}
\end{figure}

In order to generate initial transition paths to be used as starting points for aimless shooting simulations, we select the 100 configurations from the $\Sigma C_0 = 12$ umbrella sampling window and propagate them forward and backward in time with randomized initial velocities. If both forward and backward dynamics reach the same metastable basin, we perform dynamics again with new random initial velocities. In this setting, a transition path is typically achieved after less than 10 tries. Eventually, all initial contributions lead to transition pathways. The largest number of tries was 28. We report in Fig. \ref{a:lj:fig1} a histogram of the number of tries, for all 100 starting configurations (which we call the $t=0$ configurations). 

\subsection{Transition path sampling: aimless shooting}

We used the transition paths generated using brute force to initialize 100 independent aimless shooting simulations\cite{peters2006,mullen2015}. The approach is the following: starting from the $t=0$ configuration in the initial path, the system is propagated both forward and backward in time, with randomized initial velocities. Both simulations end when a metastable basin is reached. If the new path connects both basins, the configuration is stored as an "accepted" configuration, and a new starting point is obtained from the new path by selecting the configuration separated by $\pm 500 \delta t$ from the $t=0$ configuration. If the new path does not connect both basins, the configuration is stored as "rejected", and the selection strategy is applied again to the former path. The aimless shooting selection step ($500 \delta t$) has been adjusted to roughly match a $35\%$ acceptance ratio, which represents a good balance between sampling quality (a small selection step leads to highly correlated configurations), and efficiency (there is a large enough number of accepted paths). For each initial path, we perform 2200 aimless shooting iterations. Finally, 5 "accepted" and 5 "rejected" configurations per aimless shooting simulation are selected, evenly spaced across both datasets. This leads to a final dataset of 500 "accepted" and 500 "rejected" configurations. 

\subsection{Numerical estimation of the committor}

\begin{figure}[h!]
\centering
\includegraphics[width=.5\textwidth]{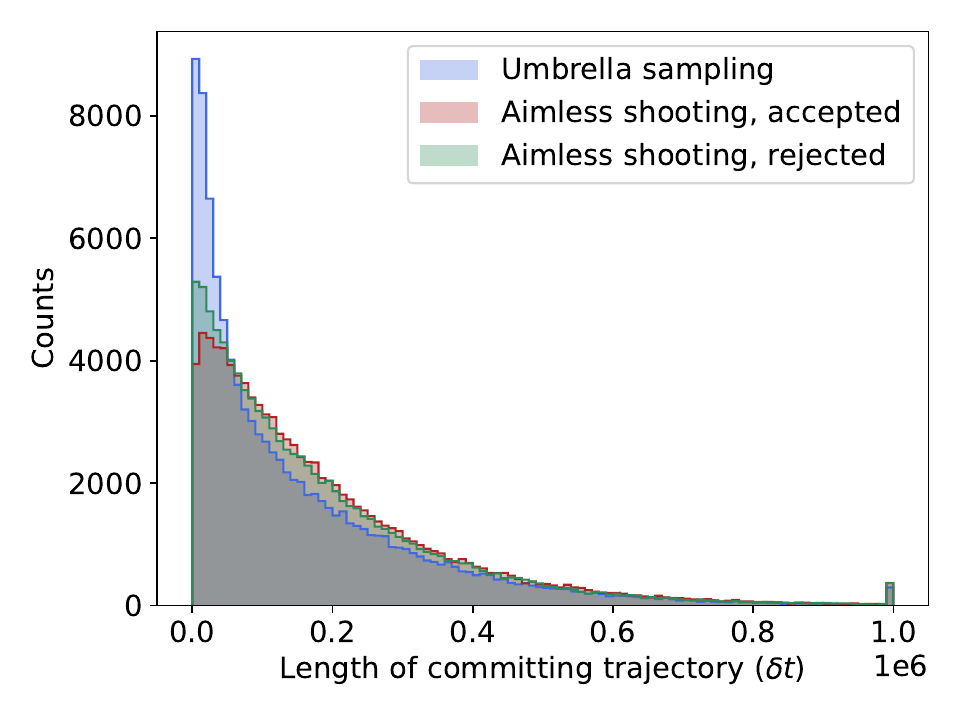}
\caption{Committing trajectory lengths, in $\delta t$. There is a peak at $10^6 \delta t$ since this is the maximum trajectory length we allow. }
\label{a:lj:fig2}
\end{figure}

To compute the committor of each configuration from the sampled configurations, we launch 200 dynamics with randomized initial velocities, which end once a basin is reached, or once the trajectory reaches $10^6 \delta t$. In this case, which represents about $0.35 \%$ of all trajectories, it is discarded. Histograms of trajectory lengths are displayed on Fig. \ref{a:lj:fig2}. The committor is then evaluated based on the number of outcomes. Overall, this required about $5.2 \cdot 10^{10}$ molecular dynamics time steps, which highlights the cost of evaluating the committor for systems showing slow committment kinetics. 

\subsection{A model selection strategy for data points and dimensionality}\label{a:modsel}

\begin{figure}[h!]
\centering
\includegraphics[width=\textwidth]{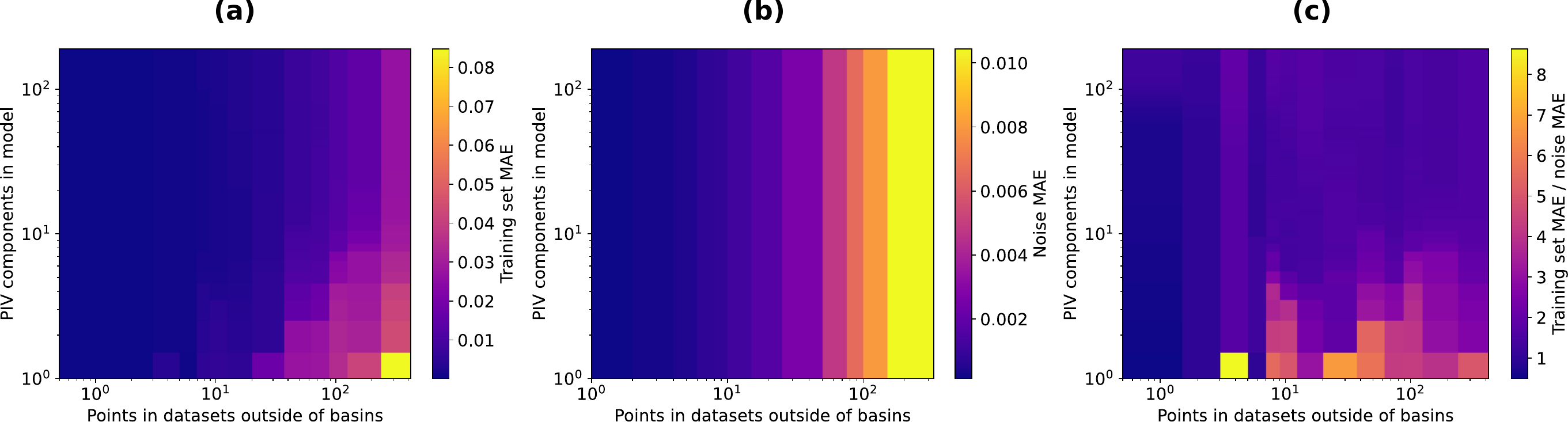}
\caption{Model selection strategy. (a) Training set MAE, (b) noise MAE, and (c) ratio of the two quantities as a function of the number of reference points included outside of basins, and of the number of PIV components included in the KRRCV mode. }
\label{a:lj:fig3}
\end{figure}

When trying to minimize the amount of committor evaluations, the map reported in Fig. \ref{fig6}(b) is generally not available, since the metric used for discrimination is the performance on the full test set. What is directly available is the performance on a small test set, or on the small training set (Fig. \ref{a:lj:fig3}(a)). The latter quantity does not allow to select the appropriate minimal number of data points outside of basins, since it will be minimal for the smallest datasets, resulting in significant overfitting. However, one can also estimate the noise MAE as a function of the dataset distribution (Fig. \ref{a:lj:fig3}(b)), following Appendix \ref{a:bd}, which decreases as a function of the number of points in the datasets. This quantity can be used as a baseline to prevent overfitting: in Fig. \ref{a:lj:fig3}(c), we plot the training set MAE divided by the noise MAE. When this quantity is smaller or close to one, overfitting is significant. When it is large (\textit{e.g.} at low dimensionality), the model performs poorly even on the training set. When it reaches an intermediate value ($\approx 2-3$), it seems to lead to appropriate models, balancing accuracy and overfitting. As a strategy, we therefore recommend to optimize models with a small number of data points, progressively adding more points until this quantity reaches $\approx 2-3$. 

\section{Uncertainty on numerical estimates of the committor probability: mean absolute error}\label{a:bd}
Since the committor must be estimated numerically, there is a numerical uncertainty associated that leads to a lower bound on the MAE, \textit{i.e.} even if the KRR would correlate perfect with the comittor, we would get a finite value of the MAE that can be estimated as follows. The numerical estimation of the committor probability $p_B$ through $N$ repeated and independent trial molecular dynamics simulations is a Bernoulli process. The number of successes $k = N \cdot p_B$, \textit{i.e.} the number of trajectories committing to basin $B$ for $N$ trials, therefore follows a binomial distribution. The mean absolute error (MAE) of a binomially-distributed random variable has a closed-form expression identified by de Moivre\cite{demoivre1756,diaconis1991}: 

\begin{equation}\label{a:bd:mae}
\text{MAE}(p_B, N) = \mathbb{E} \left| p_B - \mathbb{E} p_B \right| = \frac{1}{N} 2k \left(1 - p_B\right) {N\choose k} b \left(  k, N, p_B \right), 
\end{equation}

where $b \left(  k, N, p_B \right)$ is the probability mass function of the binomial distribution: 

\begin{equation}\label{a:bd:pmf}
b \left(  k, N, p_B \right) = p_B^k \left( 1 - p_B \right)^{N-k}.
\end{equation}

We can therefore compute the MAE as a function of the committor probability, and of the number of trials. For the LiF association in water, since datasets are uniform in committor values, we can estimate the MAE on data as: 

\begin{equation}
    \text{MAE}(N) = \int_0^1 \text{MAE}(p_B, N) \mathrm{d} p_B.
\end{equation}

\begin{figure}[h!]
\centering
\includegraphics[width=.5\textwidth]{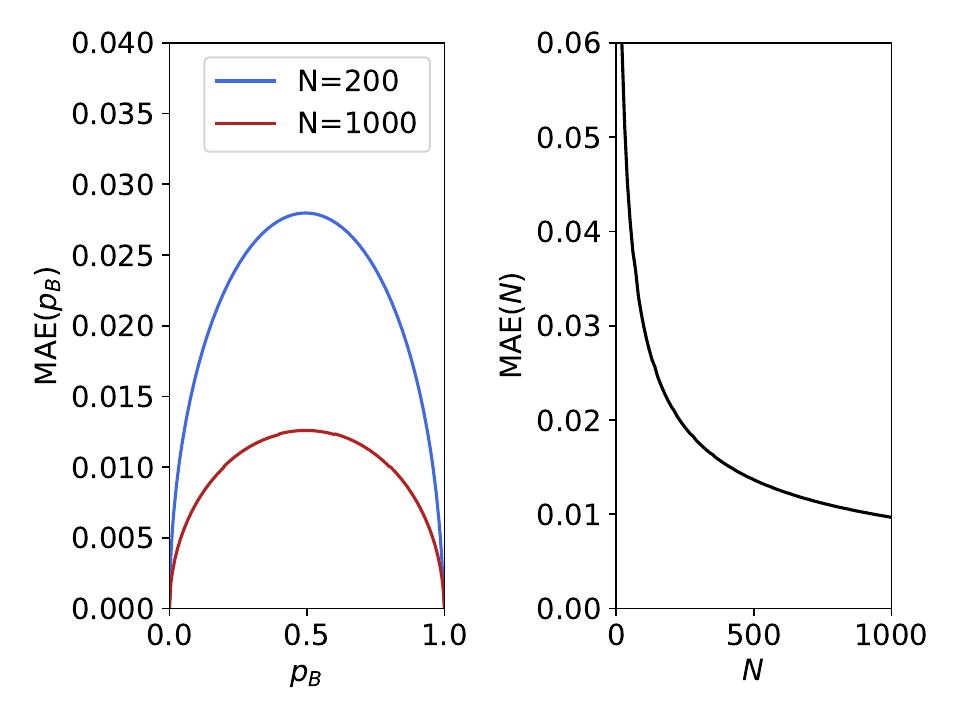}
\caption{Numerical estimation of the committor probability. Left: MAE as a function of committor value, right: MAE as a function of the number of trials.}
\label{a:bd:fig1}
\end{figure}

The MAE on data for homogeneous datasets is reported in Fig. \ref{a:bd:fig1}, with the MAE dependence on both $p_B$ and $N$. For the LiF association in water, $N=1000$ and $\text{MAE} \approx 0.010$. The Lennard-Jones datasets being heterogeneously distributed along $p_B$, we use the actual test set distribution to evaluate MAE($N$), \textit{i.e.}, for a dataset with $M$ data points:

\begin{equation}
    \text{MAE}(N) = \frac{1}{M} \sum_{i=1}^{M} \text{MAE}(p_B^i, N)
\end{equation}

For $N = 200$, $\text{MAE} \approx 0.017$. A homogeneously-distributed dataset would lead to $\text{MAE} \approx 0.021$; the -- small -- reduction is due to the larger amount of basin configurations, for which the error is minimal. 

\section{LiF association in water: computational details and additional information}\label{a:lif}

\subsection{System generation, initial relaxation}

We generate 100 initial configurations using a Monte Carlo algorithm implemented in the packmol program~\cite{martinez2009}, with one cation, one anion, and 160 water molecules, in a cubic box with 16.90 \AA-long edges. The atomic velocities are initialized by drawing from the Maxwell-Boltzmann distribution at $T = 300 \text{K}$. We relax these configurations in the $npT$ ensemble at $T = 300 \text{K}$, $p = 1 \text{atm}$, for 5 ns. The box size is subsequently fixed at the average equilibrium value obtained, 16.83 \AA. 

\subsection{Unbiased free energy estimates}

From the previous, relaxed geometries, we perform 100 independent simulations in the $nVT$ ensemble with randomized initial velocities, 1 ns of equilibration, and 20 ns for sampling, which amounts to a total of 2 $\mu$s of dynamics. The interionic distance ($r$) is computed every 10 fs; this dataset is subsequently used to compute the free energy profile along $r$ by binning. Uncertainty estimates are obtained by computing 95\% confidence intervals over the distribution made of the 100 estimates. 

\subsection{Umbrella sampling simulations}

We perform a single umbrella sampling simulation by constraining the interionic distance at $r_0 = 2.636$ \AA, using a harmonic biasing potential with a force constant set to 5000 kcal/mol/\AA, for 50 ns. Configurations are sampled every 10 ps; we therefore obtain a dataset of 5000 configurations matching the constraint on $r$. 

\subsection{Transition path sampling: brute force}

In order to generate initial transition paths to be used as starting points for aimless shooting simulations, we randomly select 200 configurations with $0.3 \leq p(\text{B}|\mathbf{X}) \leq 0.7$ from the umbrella sampling dataset, and propagate them forward and backward in time with randomized initial velocities. If both forward and backward dynamics reach the same metastable basin, we perform dynamics again with new random initial velocities. Transition paths are achieved with less than 5 tries for all configurations; as shown in Fig. \ref{fig12}(a), $p(\text{TP}|\mathbf{X})$ is high for transition state configurations. 

\subsection{Transition path sampling: aimless shooting}

Starting from the previously generated transition paths, we perform 200 independent aimless shooting simulations of $10^4$ iterations, with a selection step of 10 fs, leading to an average acceptance ratio of 42\%. Finally, we downsample the list of sampled structures by a factor of 100, leading to a dataset of 8551 configurations. 

\subsection{Numerical estimation of the committor and of the transition path probability}

We obtain numerical estimates of the committor by launching 1000 independent unbiased dynamics from each configuration. We also perform backward dynamics to estimate $p(\text{TP}|\mathbf{X})$; these backward dynamics statistics are however discarded when estimating $p(\text{B}|\mathbf{X})$. 

\subsection{Committor distribution at the critical interionic distance from unbiased molecular dynamics}\label{a:pbdumd}

\begin{figure}[h!]
\centering
\includegraphics[width=0.5\textwidth]{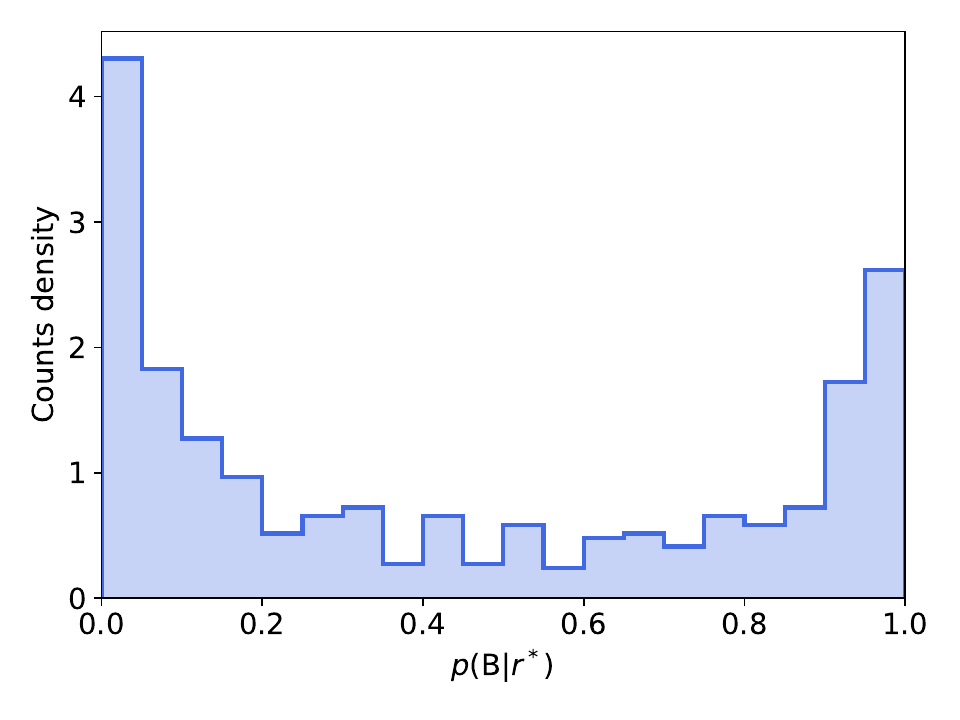}
\caption{Distribution of committor values for configurations at the putative transition state ensemble of $r$ sampled from unbiased simulations. }
\label{a:lif:fig1}
\end{figure}

We investigate the shape of the committor distribution for configurations matching $r \approx r^{*}$ sampled from unbiased molecular dynamics. We perform 100 independent simulations of 10 ns, amounting to a total sampling time of 1 $\mu$s. Every 10 fs, configurations for which $r \in [2.55, 2.70]$~\AA\ are saved. We obtain 2706 configurations, out of which 581 are separated by at least 1 ps, for which we compute the committor (using 500 velocity initializations). The distribution, whose shape matches the one of the umbrella sampling distribution, is displayed in Fig. \ref{a:lif:fig1}. 

\bibliography{biblio}

\end{document}